\documentclass{aa}
\usepackage{subfigure}
\usepackage{graphicx}
\usepackage{txfonts}
\usepackage{upgreek}
\usepackage{natbib}
\usepackage{xcolor}
\usepackage[colorlinks=true,linkcolor=blue,citecolor=blue,urlcolor=blue]{hyperref}

\def\as {\ifmmode {\rlap.}$\,$''$\,$\! \else ${\rlap.}$\,$''$\,$\!$\fi}

\begin{document} 

\title{PRODIGE - envelope to disk with NOEMA.}

\subtitle{IV. An infalling gas bridge surrounding two Class 0/I systems in L1448N}

\author{C. Gieser
	\inst{1}
	\and
	J.~E. Pineda\inst{1}
	\and
	D.~M. Segura-Cox\inst{2, 1}
	\and
	P. Caselli\inst{1}
	\and
	M.~T. Valdivia-Mena\inst{1}
	\and
	M.~J. Maureira\inst{1}
	\and
	T.~H. Hsieh\inst{1}
	\and
	L.~A. Busch\inst{1}
	\and
	L. Bouscasse\inst{3}
	\and
	A. Lopez-Sepulcre\inst{3,4}
	\and
	R. Neri\inst{3}
	\and
	M. Kuffmeier\inst{5}
	\and
	Th. Henning\inst{6}
	\and
	D. Semenov\inst{6}
	\and
	N. Cunningham\inst{7}
	\and
	I. Jimenez-Serra\inst{8}
	}

	\institute{Max Planck Institute for Extraterrestrial Physics, Gießenbachstraße 1, 85749 Garching bei M\"unchen, Germany\\
	\email{gieser@mpe.mpg.de}
	\and
	Department of Astronomy, The University of Texas at Austin, 2515 Speedway, Austin, TX 78712, USA
	\and
	Institut de Radioastronomie Millimétrique (IRAM), 300 rue de la Piscine, F-38406, Saint-Martin d’Hères, France
	\and
	IPAG, Universit\'{e} Grenoble Alpes, CNRS, F-38000 Grenoble, France
	\and
	Centre for Star and Planet Formation, Niels Bohr Institute, University of Copenhagen, Øster Voldgade 5-7, DK-1350 Copenhagen, Denmark
	\and
	Max-Planck-Institut für Astronomie, Königstuhl 17, D-69117 Heidelberg, Germany
	\and
	SKA Observatory, Jodrell Bank, Lower Withington, Macclesfield SK11 9FT, United Kingdom
	\and
	Centro de Astrobiología (CAB), CSIC-INTA, Ctra.deTorrejón a Ajalvir km 4, 28806, Torrejón de Ardoz, Spain
	}

	\date{Received x; accepted x}

	\abstract
	{The formation of stars has been subject to extensive studies in the past decades from molecular cloud to protoplanetary disk scales. It is still not fully understood how the surrounding material in a protostellar system, that often shows asymmetric structures with complex kinematic properties, feeds the central protostar(s) and their disk(s).}
	{We study the spatial morphology and kinematic properties of the molecular gas surrounding the IRS3A and IRS3B protostellar systems in the L1448N region located in the Perseus molecular cloud.}
	{We present 1\,mm Northern Extended Millimeter Array (NOEMA) observations of the large program ``PROtostars \& DIsks: Global Evolution'' (PRODIGE) and analyze the kinematic properties of molecular lines. Given the complexity of the spectral profiles, the lines are fitted with up to three Gaussian velocity components. The clustering algorithm ``Density-Based Spatial Clustering of Applications with Noise'' (\texttt{DBSCAN}) is used to disentangle the velocity components into the underlying physical structure.}
	{We discover an extended gas bridge ($\approx$3000\,au) surrounding both the IRS3A and IRS3B systems in six molecular line tracers (C$^{18}$O, SO, DCN, H$_{2}$CO, HC$_{3}$N, and CH$_{3}$OH). This gas bridge is oriented along the northeast-southwest direction and shows clear velocity gradients on the order of 100\,km\,s$^{-1}$\,pc$^{-1}$ towards the IRS3A system. We find that the observed velocity profile is consistent with analytical streamline models of gravitational infall towards IRS3A. The high-velocity C$^{18}$O ($2-1$) emission towards IRS3A indicates a protostellar mass of $\sim$1.2\,$M_\odot$.}
	{While high angular resolution continuum data often show IRS3A and IRS3B in isolation, molecular gas observations reveal that these systems are still embedded within a large-scale mass reservoir with a complex spatial morphology as well as velocity profiles. The kinematic properties of the extended gas bridge are consistent with gravitational infall toward the IRS3A protostar.}
	

	\keywords{Stars: formation -- Stars: protostars -- ISM: kinematics and dynamics -- ISM: individual objects: LDN 1448N}

	\maketitle

\section{Introduction}\label{sec:intro}

\begin{figure*}[!htb]
\centering
\includegraphics[]{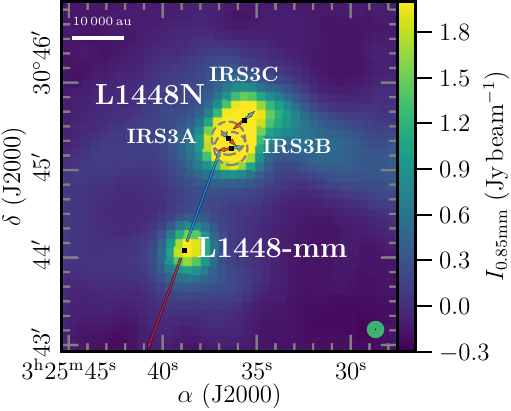}
\includegraphics[]{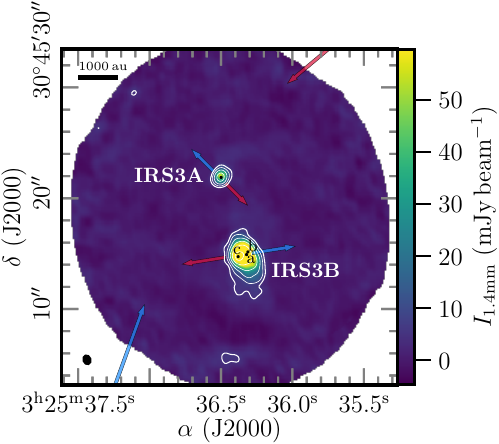}
\caption{Continuum images toward L1448N. The left panel shows in color the JCMT 850\,$\upmu$m emission taken from the COMPLETE survey \citep[][]{Ridge2006, Kirk2006}. The grey dashed circles are the primary beam (22\farcs8) of the two NOEMA pointings of the PRODIGE observations. The protostellar systems of L1448N (IRS3A, IRS3B, and IRS3C), as well as the nearby L1448-mm system are marked by black squares. The beam of JCMT and NOEMA observations is shown in the bottom right corner in green and black, respectively. In the right panel the 1.4\,mm continuum image of the PRODIGE observations is presented in color and white contours. Contour levels are 5, 10, 20, 40, 80, 160$\times \sigma_\mathrm{cont}$ ($\sigma_\mathrm{cont}$=0.94\,mJy\,beam$^{-1}$). The black circles mark the positions of individual protostars taken from the VANDAM survey \citep{Tobin2016}. The synthesized beam of the NOEMA data is shown in the bottom left corner. In both panels scale bars are marked in the top left corner and bipolar outflow orientations are highlighted by red and blue arrows (Sect. \ref{sec:outflow}).}
\label{fig:continuum}
\end{figure*}

	Star formation is a hierarchical process from (giant) filamentary molecular clouds of up to several hundreds of parsec in size \citep{Ragan2014} down to protostellar cores with sizes $<$0.1\,pc harboring protostars \citep{Pineda2023}. The most embedded phase of low-mass star formation (with final stellar masses $M_\star$ lower than $\approx$2\,$M_\odot$) is the Class 0/I stage during which the protostellar systems are still embedded within their natal core envelope \citep{Evans2009,Dunham2014}. 
	
	Interferometric observations in the millimeter (mm) wavelength range with, for example, the Northern Extended Millimeter Array (NOEMA), Submillimeter Array (SMA), and the Atacama Large Millimeter/submillimeter Array (ALMA), allow us to study nearby star-forming regions on scales of a few 1000\,au down to several au tracing the cold gas and dust. High angular resolution ($\lesssim$0\as3) surveys revealed the presence of structured planet-forming disks with a diversity of physical and chemical properties \citep[e.g.,][]{Fedele2017, Andrews2018, SeguraCox2020, Oberg2021, Ohashi2023, Miotello2023}. Stars are often associated in multiple systems \citep[e.g.,][]{Offner2023}, where in the early evolutionary phases the multiplicity fraction can be even higher \citep[e.g.,][]{Tobin2016}. 
	
	Extended emission is often filtered out in interferometric observations, i.e. the surrounding envelope material. Since the envelope can be the main mass reservoir, it is important to study star-forming regions on intermediate scales of a few 100\,au to fill the gap between single dish and interferometric observations. This regime is crucial to connect the large-scale cloud/filament/core properties with the characteristics of the protostar(s) and protoplanetary disks. In particular, it is important to trace the gas flows from filament and core scales down to the protostars and disks. Protostellar envelopes have asymmetric and complex structures \citep[e.g.,][]{Tobin2010,Tobin2011,Tobin2012,Maureira2017}. The kinematic properties of the molecular gas surrounding protostars reveal asymmetric gas flows with infall motions - sometimes referred to as streamers - on scales of a few 100\,au up to several 1\,000\,au \citep[e.g.,][]{Yen2014, Yen2017, Pineda2020, Ginski2021, Garufi2022, ValdiviaMena2022, Hsieh2023, FernandezLopez2023}. These streamers can be traced by a variety of species, such as CS, DCN, HCO$^{+}$, H$_{2}$CO, HC$_{3}$N, CH$_{3}$CN, and CH$_{3}$OH and their chemical properties as well as their connection to the natal environment is still active research. While early protostellar gravitational collapse models assumed a spherically symmetric envelope forming a single protostar \citep[e.g.,][]{Larson1969,Shu1977}, more sophisticated models and simulations have been developed over the recent years in which multiple systems can form as well \citep[][and references within]{Kuffmeier2024}. Analytic infall models assuming spherical symmetry can describe the kinematics of the asymmetric gas by tracing the kinematic profile of material along streamlines \citep{Pineda2020, ValdiviaMena2022}.
	
	In this work, we study the kinematic properties of gas flows in the L1448N region, located in the Perseus molecular cloud at a distance of 288\,pc \citep[estimated by combining Gaia astrometry, stellar photometry, and CO maps,][]{Zucker2018}, and how these flows connect to the embedded protostellar systems IRS3A and IRS3B. The left panel in Fig. \ref{fig:continuum} presents an overview of the cold dust distribution surrounding the L1448N region traced by 850\,$\upmu$m continuum emission taken from the Coordinated Molecular Probe Line Extinction and Thermal Emission (COMPLETE) survey \citep[][]{Ridge2006} with the James Clerk Maxwell Telescope (JCMT). Toward L1448N a dense dust core harbors three protostellar systems: IRS3A, IRS3B, and IRS3C. The Class 0 protostar L1448-mm is located at a projected distance of $\approx$25\,000\,au toward the southeast. 
	
	Protostars in the Perseus molecular cloud have been subject to extensive studies in the past tracing both the large and small scale environment \citep[e.g.,][]{Jorgensen2007,Jorgensen2009}. Within the Mass Assembly of Stellar Systems and their Evolution (MASSES) survey, envelope masses of 0.10\,$M_\odot$ and 0.34\,$M_\odot$ and outflow position angles (PA) of 218$^\circ$ and 122$^\circ$ are derived for IRS3A and IRS3B, respectively \citep{Lee2016}. The bolometric luminosity of IRS3A and IRS3B is estimated to be $9.2\pm1.3$ and $8.3\pm0.8$\,$L_\odot$, respectively \citep{Tobin2016}. L1448N hosts at least 6 protostars: The VLA Nascent Disk and Multiplicity Survey of Perseus Protostars (VANDAM) revealed that IRS3A is a single protostar, IRS3C is a binary system, and IRS3B is a triple system (IRS3B-a, IRS3B-b, and IRS3B-c hereafter, marked in the right panel in Fig. \ref{fig:continuum}) \citep{Tobin2016}. While IRS3B-c is dominating the mm emission, the close binary IRS3B-a and IRS3B-b dominate in mass \citep{Tobin2016b}. About 1 arcmin away from L1448N, L1448-mm, also known as L1448C or Per-emb-26, is also part of a multiple system with the nearby L1448C(S) system, also known as Per-emb-42 \citep{Jorgensen2006, Tobin2016}.
	
	Several surveys have targeted these protostars to account for the chemical complexity in these regions. The Continuum And Lines in Young ProtoStellar Objects (CALYPSO) do not reveal any secure detections of complex organic molecules in IRS3A and IRS3B, except for a tentative detection of methanol (CH$_{3}$OH) and methyl formate (CH$_{3}$OCHO) \citep{DeSimone2017, Belloche2020}. In the Perseus ALMA Chemistry Survey (PEACHES) detections of CH$_{3}$OH, CH$_{3}$CN, CH$_{3}$OCHO are reported toward IRS3A and IRS3B-c and none of them toward IRS3B-a/b \citep{Yang2021}. Most notably, these authors find that while for most of their sample the emission of CH$_{3}$OH is compact surrounding the protostars, for IRS3A the emission is extended.

	The orientations of known bipolar outflows are marked in Fig. \ref{fig:continuum}. A large-scale outflow is launched by L1448-mm \citep[e.g.,][]{Bally1993, Bachiller1995, Barsony1998,JimenezSerra2011} with the blueshifted lobe directed toward L1448N. \citet{ToledanoJuarez2023} find that the outflows in L1448-mm with a northwest-southeast direction interacts with the close-by L1448-C(S) outflow. The redshifted lobes of both outflows are colliding with each other, however, these authors find no evidence for triggered star formation. A mid-infrared (MIR) study by \citet{OLinger2006} revealed that only L1448-mm and L1448 IRS3A are bright in the mid-infrared (from 12.5\,$\upmu$m to 24.5\,$\upmu$m). By modeling the spectral energy distribution (SED) from MIR to mm wavelengths, these authors find for both IRS3A and L1448-mm systems two temperature components at $\approx$20\,K and $\approx$90\,K. These authors suggested that IRS3A is close to being a Class I system by comparing the envelope mass $M_\mathrm{env}=0.29$\,$M_\odot$ to Class 0 and I systems from the literature. These authors also concluded that the non-detection of the IRS3B system in the MIR might be related to the IRS3A envelope being disrupted by the L1448-mm outflow, while IRS3B is still embedded within its natal envelope.
	
	\citet{Tobin2016b} and \citet{Reynolds2021} present high angular resolution observations ($\lesssim$0\farcs3) with ALMA revealing spiral structures in the IRS3B disk (with a size of 497\,au $\times$ 351\,au) as well as potentially for IRS3A. \citet{Reynolds2021} claim that the triple system in IRS3B formed via disk fragmentation - IRS3B-a and IRS3B-b could have formed first and the tertiary component IRS3B-c recently formed via gravitational instability. Contrary to that, these authors find that the disk of IRS3A, with a size of 197\,au $\times$ 72\,au, is gravitationally stable. At a spatial resolution of 8\,au, ALMA reveals that the IRS3A disk structure is a ring and that potentially a fourth protostar is embedded in the IRS3B disk \citep{Reynolds2024}.
	
\setlength{\tabcolsep}{5pt}
\begin{table*}[!htb]
\caption{Properties of molecular lines covered by the NOEMA PRODIGE observations.}
\label{tab:obs}
\centering
\renewcommand{\arraystretch}{1.1}
\begin{tabular}{lrr rrr rrrr}
\hline\hline
Transition & Frequency & Upper energy & Einstein $A$ & & & Channel & Line & \multicolumn{2}{c}{Moment 0} \\
 & & level & coefficient & \multicolumn{2}{c}{Synthesized Beam} & width & noise & \multicolumn{2}{c}{integration range} \\ \cline{5-6} \cline{9-10}
 & $\nu$ & $E_\mathrm{u}$/$k_\mathrm{B}$ & log $A_\mathrm{ul}$ & $\theta_\mathrm{maj}\times\theta_\mathrm{min}$ & PA & $\delta \varv$ & $\sigma_\mathrm{line}$ & $\varv_\mathrm{low}$ & $\varv_\mathrm{upp}$ \\
 & (GHz) & (K) & (log s$^{-1}$) & ($'' \times ''$) & ($^\circ$) & (km\,s$^{-1}$) & (K) & (km\,s$^{-1}$) & (km\,s$^{-1}$)\\
 \hline
SiO\,($5-4$) & 217.104980 & 31.3 & $-3.3$ & 1.24$\times$0.88 & 15 & 0.1 & 0.19 & $\ldots$ & $\ldots$\\ 
 & & & & 1.24$\times$0.88 & 15 & 3.0 & 0.037 & $-50.0$ & $60.0$\\ 
DCN\,($3-2$) & 217.238538 & 20.9 & $-3.3$ & 1.24$\times$0.88 & 15 & 0.1 & 0.22 & $1.0$ & $10.0$\\ 
H$_{2}$CO\,($3_{0,3}-2_{0,2}$) & 218.222192 & 21.0 & $-3.6$ & 1.23$\times$0.88 & 15 & 0.1 & 0.25 & $1.0$ & $11.0$\\ 
HC$_{3}$N\,($24-23$) & 218.324723 & 131.0 & $-3.1$ & 1.23$\times$0.88 & 15 & 0.1 & 0.22 & $1.5$ & $8.0$\\ 
CH$_{3}$OH\,($4_{2,3}-3_{1,2}E$) & 218.440063 & 45.5 & $-4.3$ & 1.23$\times$0.88 & 15 & 0.1 & 0.22 & $2.5$ & $11.0$\\ 
OCS\,($18-17$) & 218.903356 & 99.8 & $-4.5$ & 1.23$\times$0.88 & 13 & 0.1 & 0.22 & $1.0$ & $11.0$\\ 
C$^{18}$O\,($2-1$) & 219.560357 & 15.8 & $-6.2$ & 1.22$\times$0.87 & 13 & 0.1 & 0.21 & $1.0$ & $11.0$\\ 
SO\,($6_{5}-5_{4}$) & 219.949442 & 35.0 & $-3.9$ & 1.22$\times$0.87 & 13 & 0.1 & 0.30 & $-1.5$ & $12.0$\\ 
CO\,($2-1$) & 230.538000 & 16.6 & $-6.2$ & 1.18$\times$0.83 & 14 & 0.1 & 0.35 & $\ldots$ & $\ldots$\\ 
 & & & & 1.18$\times$0.83 & 14 & 3.0 & 0.038 & $-50.0$ & $55.0$\\ 
\hline
\end{tabular}
\tablefoot{The transition properties ($\nu$, $E_\mathrm{u}$/$k_\mathrm{B}$, and $A_\mathrm{ul}$) are taken from the CDMS \citep{CDMS} and JPL \citep{JPL} databases. The noise per channel in the line data, $\sigma_\mathrm{line}$, is estimated in the central area of the mosaic as the standard deviation in line-free channels. For SiO and CO, we use in this work both the wideband and narrowband data with channel widths of 3.0\,km\,s$^{-1}$ and 0.1\,km\,s$^{-1}$, respectively. The last two columns show the velocity ranges, $\varv_\mathrm{low}$ and $\varv_\mathrm{upp}$, used to compute the line integrated intensity (moment 0) maps presented in Fig. \ref{fig:moment0}.}
\end{table*}

	In summary, the protostellar systems in L1448N have been substantially studied from IR to cm wavelengths targeting both the large-scale structures ($>$ few thousand of au) and the smallest scales down to the disks and protostars ($<$ few tens of au). However, important aspects to understand the formation of these low-mass protostellar systems have not yet been addressed: How does the material flow onto the protostellar systems and is there a connection to the larger-scale filament (Fig. \ref{fig:continuum} left panel) or in between? What is the impact of the blueshifted outflow launched by L1448-mm onto the L1448N region and could it have triggered the formation of the systems in L1448N as it has been suggested in the literature \citep[e.g.,][]{Barsony1998, OLinger2006}? 
	
	A survey targeting such scales (300\,au to few 1000 au) is the NOEMA large program PROtostars \& DIsks: Global Evolution (PRODIGE). The PolyFiX correlator of NOEMA tuned to 1.4\,mm wavelengths provides a broad bandwidth of 16\,GHz that allows to simultaneously observe a plethora of molecular lines and obtain a sensitive continuum image. In total, 32 Class 0/I systems toward the Perseus molecular cloud are covered by the ongoing PRODIGE project, resolving scales down to $\approx$300\,au with an angular resolution of $\approx$1$''$. The first PRODIGE results of the program already revealed infalling streamers in H$_{2}$CO in Per-emb-50 \citep{ValdiviaMena2022} and in DCN emission in the hot corino SVS13A \citep{Hsieh2023, Hsieh2024}. The second part of PRODIGE targeting eight Class II disks in Taurus reveal compact and extended disks in $^{12}$CO, $^{13}$CO and C$^{18}$O (2-1) line emission \citep{Semenov2024} and by modeling these species, disk gas masses between 0.005 and 0.04\,$M_\odot$ are determined \citep{Franceschi2024}. The PRODIGE program covers two NOEMA pointings toward IRS3A and IRS3B, as sketched by the grey dashed circles in Fig. \ref{fig:continuum}.
	
	In this work, we investigate the kinematic properties of the cold molecular gas in the L1448N region using molecular line observations with NOEMA toward the IRS3A and IRS3B systems and how these may be affected by the nearby protostellar systems IRS3C and L1448-mm. The paper is organized as followed: In Sect. \ref{sec:observations} we explain the data calibration procedure, including phase self-calibration, mosaic combination, and imaging of the NOEMA data. The analysis of the molecular line morphology and kinematic properties are presented in Sect. \ref{sec:results}. Our findings are discussed in Sect. \ref{sec:discussion} and our conclusions are summarized in Sect. \ref{sec:conclusions}.
	
\section{Observations}\label{sec:observations}

	The NOEMA observations were obtained within the PRODIGE project (program id: L19MB, PIs: P. Caselli and Th. Henning). One part of PRODIGE targets in total 32 Class 0/I protostars in the Perseus molecular cloud, including two NOEMA single pointing observations centered on the IRS3A and IRS3B systems in L1448N (Fig. \ref{fig:continuum}).
	
	\subsection{Single pointing observations}
	
	Both IRS3A and IRS3B were observed with a single pointing each at 1.4\,mm with a NOEMA primary beam of size 22\farcs8 (indicated by dashed circles in the left panel in Fig. \ref{fig:continuum}). The phase centers (Right Ascension $\alpha$ and Declination $\delta$ (J2000)) for the IRS3A and IRS3B pointings are 03$^h$25$^m$36.499$^s$ $+$30$^\circ$45$'$21\farcs88 and 03$^h$25$^m$36.379$^s$ $+$30$^\circ$45$'$14\farcs73, respectively. The observations in C- and D-configurations with 10 or 11 antennas in the array were taken from December 2019 to August 2021 covering baselines from 16\,m (12\,k$\uplambda$) to 370\,m (272\,k$\uplambda$). These baseline ranges roughly correspond to spatial scales of 6000\,au down to 300\,au at the distance of L1448N \citep[288\,pc,][]{Zucker2018}. 
	
	The PolyFiX correlator was set to a tuning frequency in the 1.4\,mm range, covering frequencies from 214.7\,GHz to 222.8\,GHz in the lower sideband and from 230.2\,GHz to 238.3\,GHz in the upper sideband, with a channel width of 2\,MHz ($\approx$2.6\,km\,s$^{-1}$), referred to as wideband data in the following. The wideband consists of four basebands covering $\approx$4\,GHz each. Within the wideband frequency coverage, 39 high spectral resolution (narrowband) units were placed targeting molecular lines at a higher spectral resolution of 62.5\,kHz ($\approx$0.09\,km\,s$^{-1}$).
	
	The NOEMA data were calibrated using the Continuum and Line Interferometer Calibration (CLIC) package in GILDAS\footnote{\url{https://www.iram.fr/IRAMFR/GILDAS/}}. Bright radio sources (3C\,84 or 3C\,454.3) were used to calibrate the frequency bandpass. For amplitude and phase calibration 0322+222 and 0333+321 were jointly observed every $\sim$20 minutes within each observation track. For absolute flux calibration, we used LkHA\,101, 2010+723 and/or MWC\,349. With NOEMA, flux calibration is correct at a 20\% level or better at 1.4\,mm.
	
	Further processing of the data was done with the MAPPING package of GILDAS. In each wideband baseband, a continuum-subtracted cube and a continuum image was created. The continuum was created with the \texttt{uv\_continuum} task after carefully masking channel ranges at the edges of the baseband and with line emission using \texttt{uv\_filter}. The same mask was applied for the continuum subtraction of the wideband cubes with the \texttt{uv\_baseline} task fitting a 1st order polynomial to visibilities in the continuum-only channels.
	
	Atmospheric instabilities cause a high phase noise, but given that both the IRS3A and IRS3B systems are bright sources at 1.4\,mm, the dynamic range can be significantly increased by applying phase self-calibration after the standard calibration procedure \citep[e.g.,][]{Gieser2021}. We perform phase self-calibration on the combined C- and D-configuration continuum data of each baseband separately, including all executions to obtain the highest signal-to-noise ratio (S/N), and then apply the gain solutions to the corresponding wideband and narrowband data.
	
	First, the continuum data were imaged with robust weighting (robust parameter 1), to maximize the spatial resolution, and with the Hogbom cleaning algorithm \citep{Hogbom1974}. A support mask was defined around the continuum sources. Self-calibration was performed in three loops with integration times decreasing from 300\,s, 135\,s, to 45\,s, and the number of iterations increasing at the same time (750, 1500, 2000 and 500, 750, 1500 for the IRS3A and IRS3B NOEMA pointing, respectively). Visibilities with gain solutions that have a S/N smaller than 6 are flagged. For the IRS3A and IRS3B NOEMA data, 1\% and 2\% of the visibilities are flagged in the self-calibration step, respectively.
	
	The gain solutions are applied to the wideband and narrowband data using the \texttt{uv\_cal} task where uncalibrated visibilities with S/N $<$6 are flagged. Continuum subtraction of the narrowband units was then performed by first flagging channels at the edges of the unit and channel ranges containing line emission with \texttt{uv\_filter} and then a 0th or 1st order polynomial was fitted to continuum-only channel ranges and subtracted with the \texttt{uv\_baseline} task.
	
	\subsection{Mosaic combination}
	
	Since the phase centers of the two NOEMA pointings are only separated by 7\farcs2, the self-calibrated continuum and line data can be combined into a mosaic. Since IRS3A and IRS3B have slightly different velocities (5.3\,km\,s$^{-1}$ and 4.7\,km\,s$^{-1}$, respectively), we adopt for the mosaic an intermediate source velocity of 5\,km\,s$^{-1}$. For individual line cubes, the velocity was set to the line center using the \texttt{modify} command and a subcube was extracted around the line using \texttt{uv\_extract}. To combine both pointings, the spectral axis was placed on the same grid using the \texttt{uv\_resample} task. In order to increase the S/N, all line cubes were rebinned to a common resolution of 0.1\,km\,s$^{-1}$. For the observed CO and SiO transitions, the width of the narrowband units were not sufficient to cover the full extent of the line emission due to broad outflow line wings. We thus also extracted wideband cubes for those two lines rebinned to a channel width of 3.0\,km\,s$^{-1}$. For both the continuum and line data, the two NOEMA pointings were then combined into a mosaic using the \texttt{uv\_mosaic} task. The phase center was set as the center between the two NOEMA pointings with $\alpha$=03$^h$25$^m$36.439$^s$ and $\delta$=$+$30$^\circ$45$'$18$\farcs$304 (J2000).

	The mosaic continuum data was cleaned with robust weighting (robust parameter of 1), with the Hogbom algorithm and a support mask down to 3$\times$ the theoretical noise. A manually defined support mask covers both the IRS3A and IRS3B continuum sources. The mosaic line data were cleaned with natural weighting to increase the line signal-to-noise ratio and the Hogbom algorithm down to the theoretical noise. For the DCN ($3-2$), HC$_{3}$N ($24-23$), and CH$_{3}$OH ($4_{2,3}-3_{1,2}E$) line cubes a support mask was defined in each channel that contained significant line emission or absorption. For both the mosaic continuum and line data, primary beam correction was applied.

	The continuum images in the four basebands show a similar morphology with the intensity increasing with frequency, as expected from cold dust emission. We thus only present in this work the continuum image created from the lower inner baseband, centered at 220.785\,GHz (1.4\,mm) as shown in the right panel in Fig. \ref{fig:continuum}. The synthesized beam has a size of 0\farcs90$\times$0\farcs73 (PA=18$^\circ$) and the continuum noise $\sigma_\mathrm{cont}$ is 0.94\,mJy\,beam$^{-1}$. 
	
	The properties of the molecular line transitions analyzed in this work are summarized in Table \ref{tab:obs}. With natural weighting applied, the synthesized beam with a size of $\approx$1\farcs2$\times$0\farcs9 is slightly larger compared to the continuum data. We estimate the line noise per channel in the central area of the mosaic in line-free channel ranges. The line noise $\sigma_\mathrm{line}$ is $\approx$0.2\,K in the high spectral resolution ($\delta \varv = 0.1$\,km\,s$^{-1}$) data and $\approx$0.04\,K in the low spectral resolution ($\delta \varv = 3.0$\,km\,s$^{-1}$) data.

\section{Results}\label{sec:results}

	The 1.4\,mm L1448N continuum map of the PRODIGE observations is presented in the right panel in Fig. \ref{fig:continuum}. Both IRS3A and IRS3B systems are clearly resolved in their cold dust emission. We also mark the positions of individual protostars detected by the VANDAM survey \citep{Tobin2016}. With an angular resolution of $\approx$1$''$ (300\,au), the three protostars in IRS3B (labeled a, b, and c) are not resolved, we can only distinguish between the outer IRS3B-c protostar and the IRS3B-ab binary system. Nevertheless, the PRODIGE observations allow us to study the spatial morphology and kinematic properties of the molecular gas on scales of a few hundred to a few thousand of au surrounding the IRS3A and IRS3B systems.

\subsection{Spatial morphology of the molecular gas}\label{sec:spatialmorphology}

\begin{figure*}[!htb]
\centering
\includegraphics[width=0.9\textwidth]{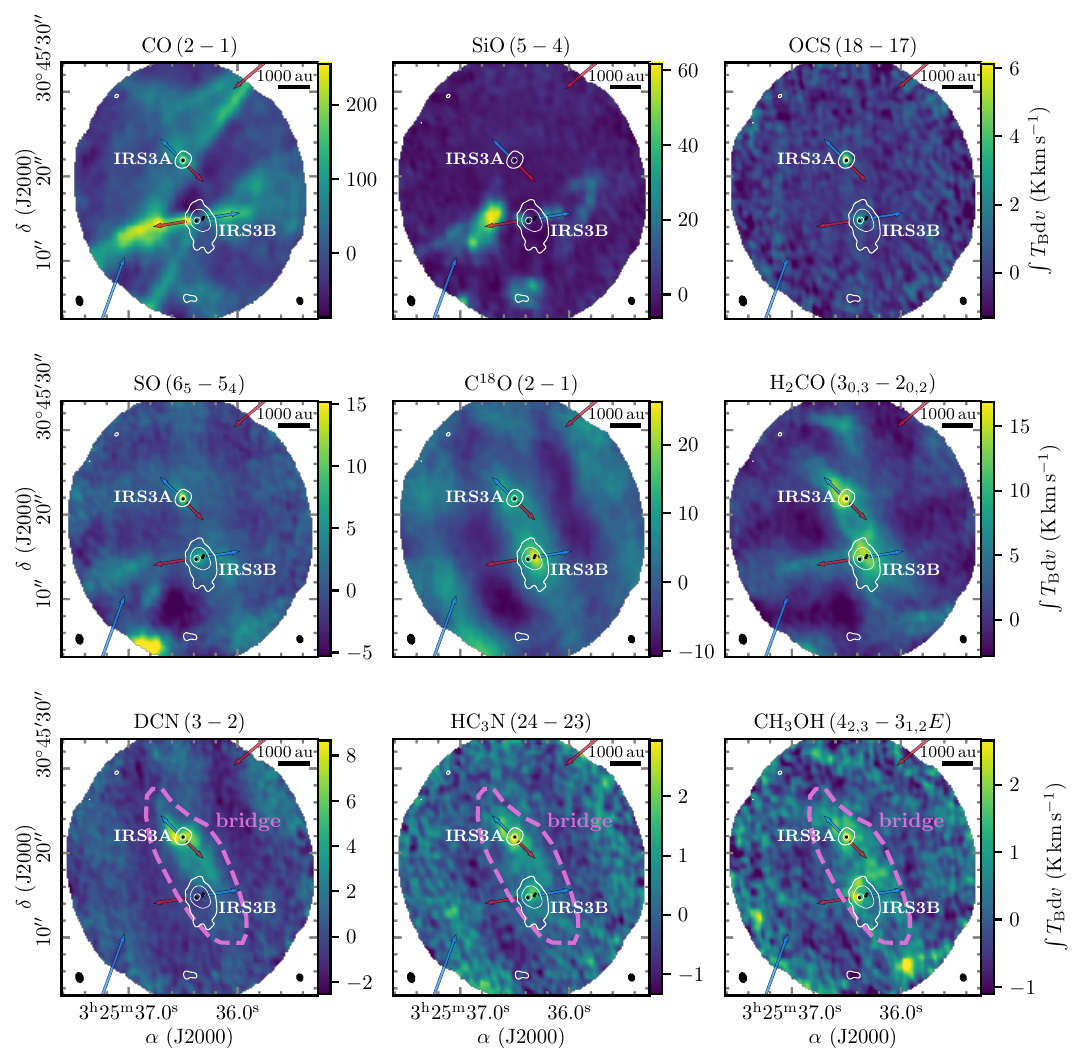}
\caption{Line integrated intensity maps of the PRODIGE observations. The line integrated intensity and the 1.4\,mm continuum is presented in color and white contours, respectively. Contour levels are 5, 40, 160$\times \sigma_\mathrm{cont}$ ($\sigma_\mathrm{cont}$=0.94\,mJy\,beam$^{-1}$). The IRS3A and IRS3B protostellar systems toward L1448N are labeled in white. The black circles mark the positions of individual protostars taken from the VANDAM survey \citep{Tobin2016}. Bipolar outflow orientations are indicated by blue and red arrows. The synthesized beam of the line and continuum data is indicated in the bottom left and bottom right corner, respectively. A scale bar is shown in the top right corner. In the bottom row panels, the extent of bridge structure is indicated by the dashed pink polygon. Spectra toward the continuum peak position of IRS3A and IRS3B are shown in Fig. \ref{fig:avg_spectra} for all transitions.}
\label{fig:moment0}
\end{figure*}

	With the PRODIGE data, common abundant molecular species are detected in the L1448N region. The detected lines analyzed in this work and their properties are summarized in Table \ref{tab:obs}. Unlike the hot corino source SVS13A where the PRODIGE data shows a plethora of transitions by complex organic molecules \citep{Hsieh2023, Hsieh2024}, CH$_{3}$OH is the most complex species detected with NOEMA at the sensitivity and angular resolution in L1448N. In particular, for each molecule only one transition is detected, thus it is not possible to reliably derive excitation temperatures and column densities.
	
	To investigate the spatial morphology, we compute line integrated intensity (moment 0) maps of all molecular lines listed in Table \ref{tab:obs}. For each transition, we integrate along all velocities that contain line emission: $\int_{\varv_\mathrm{low}}^{\varv_\mathrm{upp}} T_\mathrm{B}$($\varv$)d$\varv$. The last two columns in Table \ref{tab:obs} summarize the velocity ranges, $\varv_\mathrm{low}$ and $\varv_\mathrm{upp}$, used for each transition to compute the moment 0 map. Thus depending on the tracer, the velocity ranges can be larger (e.g., for CO with broad line wings due to the presence of molecular outflows) compared to lines with narrow line widths around the source velocity. For CO and SiO, that show line wings broader than the extent of the narrowband units, we use the wideband line data to compute the integrated intensity maps. The integrated intensity maps are presented in Fig. \ref{fig:moment0} for all nine molecular lines presented in this work. Figure \ref{fig:avg_spectra} shows spectra of all transitions extracted from the IRS3A and IRS3B continuum peak position.
	
	The moment 0 map of CO ($2-1$) shows extended emission within the full field-of-view (FoV). The emission traces bipolar outflows from the IRS3A and IRS3B systems, as well as from the nearby IRS3C and L1448-mm systems that are outside the FoV of the NOEMA mosaic (Fig. \ref{fig:continuum}). With a shortest baseline of 16\,m, spatial scales larger than $\approx$22$''$ (6600\,au) are filtered out during the observations. Missing flux due to this filtering is most prominent in CO with negative artifacts. SiO ($5-4$) has bright emission surrounding the east and west of the IRS3B system tracing shocks along the bipolar outflow and toward the south of IRS3B. No extended SiO emission is detected toward IRS3A above 3$\sigma_\mathrm{line}$. In Sect. \ref{sec:outflow}, we further analyze the observed outflows by studying the red- and blueshifted line wings of the CO and SiO emission.
	
	The OCS ($18-17$) line has compact emission around the IRS3A and IRS3B protostars tracing warmer gas with $E_\mathrm{u}$/$k_\mathrm{B} \approx 100$\,K. Although the angular resolution of the NOEMA observations is not sufficient to resolve the triple system within the IRS3B system, the OCS emission clearly peaks at the position of the IRS3B-c protostar located in the outer part of the disk. The integrated intensity map of SO ($6_5-5_4$) also shows bright emission peaks toward IRS3A, while for IRS3B the emission peaks in between of the protostars in the triple system. SO is also detected toward the east of IRS3B at the redshifted part of the outflow. The emission toward the southwest toward the edge of the FoV can be linked to shocked gas due to the L1448-mm outflow (further discussed in Sect. \ref{sec:outflow}). In addition, there is faint SO emission in between IRS3A and IRS3B systems.
	
	Extended emission surrounding both IRS3A and IRS3B systems are prominent in C$^{18}$O ($2-1$), H$_{2}$CO ($3_{0,3}-2_{0,2}$), DCN ($3-2$), HC$_{3}$N ($24-23$), and CH$_{3}$OH ($4_{2,3}-3_{1,2}E$) emission. We refer to this feature as the ``bridge'' in the following and its extent is highlighted by the pink polygon in the bottom panels in Fig. \ref{fig:moment0}. This bridge extends from the north of IRS3A down to the IRS3B disk, embedding both of these systems, and shows filamentary substructure. Toward the IRS3B disk, DCN is heavily absorbed against the bright continuum emission. The H$_{2}$CO moment 0 map traces, in addition to the bridge, the IRS3B outflow. Both C$^{18}$O and H$_{2}$CO integrated intensity maps show negative features due to spatial filtering, suggesting that the gas bridge might be even more extended. However, the emission of DCN, HC$_{3}$N, and CH$_{3}$OH is compact enough and does not suffer from substantial missing flux filtered by the interferometric observations.

\begin{figure*}[!htb]
\centering
\includegraphics[]{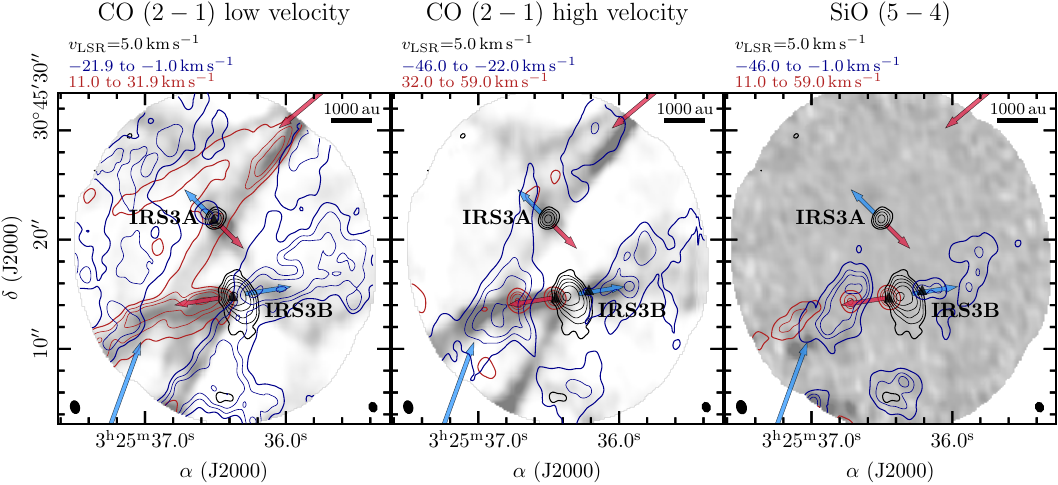}
\includegraphics[]{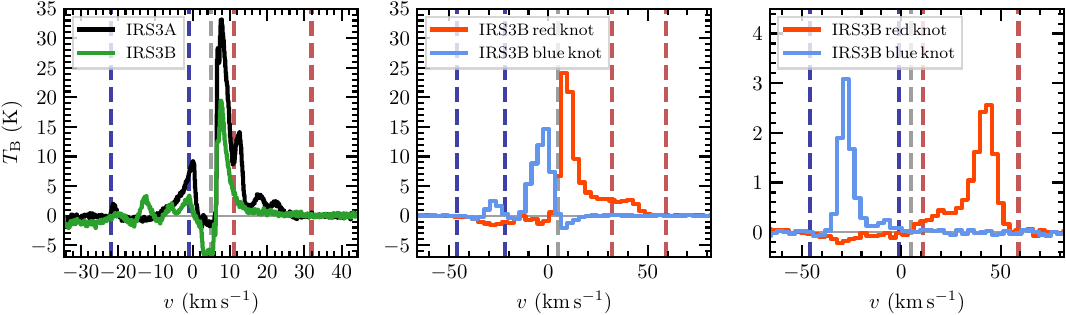}
\caption{Molecular outflows in L1448N. In the top panels, the red and blue contours show the red- and blueshifted line integrated intensities of low-velocity CO $2-1$ (left), high-velocity CO $2-1$ (center), and SiO $5-4$ (right). Contour levels are 0.1, 0.3, 0.5, 0.9$\times$ the corresponding peak integrated intensity. In color the integrated intensity around the source velocity is presented. The integration ranges are listed in each panel and indicated by red and blue dashed vertical lines in the spectra in the bottom panels. Outflow orientations are indicated by blue and red arrows. The black contours are the 1.4\,mm continuum of the PRODIGE observations. Contour levels are 5, 10, 20, 40, 80, 160$\times \sigma_\mathrm{cont}$ ($\sigma_\mathrm{cont}$=0.94\,mJy\,beam$^{-1}$). The protostellar systems of the L1448N system (IRS3A and IRS3B) are labeled in black. The synthesized beam of the line and continuum data is indicated in the bottom left and bottom right corner, respectively. A scale bar is shown in the top right corner. The bottom panels show spectra extracted from the positions indicated by the black triangles in the top panels. The grey vertical dashed line is the region velocity of L1448N ($\approx$5\,km\,s$^{-1}$).}
\label{fig:outflow}
\end{figure*}
	
	The morphology of the extended molecular emission suggests that this gas bridge is connecting the IRS3A system down to the IRS3B disk. A similar gas bridge has been found toward the Class 0 system IRAS16293$-$2422 \citep{Pineda2012, vanderWiel2019}. It was concluded by \citet{vanderWiel2019} that this quiescent bridge is remnant filamentary material from the natal protostellar envelope. The gas bridge in L1448N could be associated with the outflows or it might have a different origin. In order to investigate the properties of this bridge revealed in the PRODIGE molecular line data, we first study the outflow morphology of IRS3A and IRS3B (Sect. \ref{sec:outflow}) and in Sect. \ref{sec:kinematic} we perform a kinematic analysis of the gas bridge.

\subsection{Molecular outflows in L1448N}\label{sec:outflow}

	The identification of outflow orientations of protostars in the L1448N region in early studies was difficult \citep[e.g.,][]{Bachiller1990, Bally1993, Bachiller1995, Barsony1998} due to the close by L1448-mm outflow that has a bright and extended southeast (redshifted) - northwest (blueshifted) orientation, as indicated in Fig. \ref{fig:continuum}. Infrared observations also reveal the presence of multiple large-scale outflows originating from the L1448N region \citep[e.g.,][]{Tobin2007}.
	
	The emission of CO ($2-1$) and SiO ($5-4$) of the PRODIGE data clearly reveal bipolar outflows in L1448N (Fig. \ref{fig:moment0}). To disentangle red- and blueshifted parts of the outflows, we compute the line integrated intensity of the redshifted, central, and blueshifted velocity ranges for both transitions. For CO, we analyze both the low and high velocity components with the narrowband and wideband data, respectively. The top row in Fig. \ref{fig:outflow} shows the outflow morphology for CO (low and high velocity components in the left and middle panel, respectively) and SiO (right panel). The red- and blueshifted integrated intensities are highlighted by red and blue contours, respectively. The contour levels are 10, 30, 50, and 90\% of the peak integrated intensity. We checked for all maps that the 10\% level is larger than five times the integrated intensity noise, $\sigma_\mathrm{mom0}$. It is calculated as $\sigma_\mathrm{mom0} =\sqrt{n_\mathrm{channels}} \times \sigma_\mathrm{line} \times \delta \varv$, where $n_\mathrm{channels}$ is the number of channels used to compute the corresponding integrated intensity map, and $\sigma_\mathrm{line}$ and $\delta \varv$ are the line noise per channel and channel width, respectively (Table \ref{tab:obs}). The velocity ranges are shown in the top left corner in each panel in Fig. \ref{fig:outflow} and are highlighted by dashed vertical red and blue lines in the spectra presented in the bottom panels in Fig. \ref{fig:outflow}.
	
	The IRS3A protostar launches a northeast (blueshifted)-southwest (redshifted) outflow detected in CO ($2-1$) with no counterpart in SiO ($5-4$). The redshifted outflow lobe of IRS3A is only present close to the source velocity (up to $\varv_\mathrm{LSR}+6$\,km\,s$^{-1}$) shown in color scale and thus not visible in the red contours, which indicates that the outflow is inclined close the plane of the sky.
	
	The redshifted emission in low-velocity CO from the northwest of the FoV down to the southeast can be mistakenly identified as the outflow from IRS3A as it overlaps with the source. However, it is launched by the IRS3C system that is located in the northwest of IRS3A, but outside of the FoV of the mosaic.
	
	The IRS3B system launches southeast (redshifted)-northwest (blueshifted) outflow with shocked high-velocity bullets of CO and SiO along both outflow lobes. The bottom panels of Fig. \ref{fig:outflow} show example spectra for CO and SiO. The high-velocity SiO bullets extent up to $\pm$50\,km\,s$^{-1}$ from the source velocity.
	
	The blueshifted emission in the east and south seen in both CO and SiO are associated to the large-scale outflow of the L1448-mm protostar. In Fig. \ref{fig:outflow_largescale} we show the outflow morphology of the low velocity component of CO in L1448N as well as L1448-mm that was also observed within the PRODIGE project. The blueshifted emission blob in the northwest is also most likely associated to this outflow \citep[as suggested by][]{Girart2001}, since it has been found that the outflow lobe is asymmetric and bends towards the northwest at the location of L1448N \citep[e.g.,][]{Froebrich2002,Hirano2010,Yoshida2021}. A detailed analysis of the PRODIGE observations of the L1448-mm outflow will be presented in a different publication. The blueshifted CO emission in the low-velocity regime can also be partially caused by foreground cloud emission.
	
	All systems in L1448N show bipolar outflows. The projected size of the IRS3A outflow is small ($<$1000\,au), while for IRS3B the CO outflow traces a wide angle outflow cavity that extends well beyond the FoV. While the IRS3A outflow has no counterpart in SiO, toward the IRS3B outflow SiO traces shocked high velocity bullets that suggest energetic episodic outbursts \citep{Vorobyov2018}.
	
	In the MASSES survey, \citet{Lee2016} report an outflow PA of $218\pm10^\circ$ and $122\pm15^\circ$ for IRS3A and IRS3B, respectively. These outflow orientations are orthogonal with respect to the disks with a PA of 133$^\circ$ and 28$^\circ$, respectively \citep{Reynolds2021}. From the observed CO outflow lobes (Fig. \ref{fig:outflow}), we estimate an outflow PA of 225$^\circ$ and 99$^\circ$ for IRS3A and IRS3B, respectively. For IRS3A the outflow PA is consistent with the findings from \citet{Lee2016}. 
	
	For IRS3B, the outflow PA derived from the PRODIGE high-velocity CO data (99$^\circ$) is consistent with the high-velocity SiO bullets shown in the right panel in Fig. \ref{fig:outflow}. Toward the south-east of IRS3B, there is however a wide-angle cavity seen in CO close to the source velocity. Our results are consistent with the high velocity SiO bullets launched by IRS3B-c revealed by ALMA \citep{Reynolds2021}, while the wide angle outflow with a PA of $120-130^\circ$ \citep{Lee2016,Dunham2024} is associated with the compact IRS3B-ab system.

\subsection{Source velocity and mass estimates}\label{sec:PV}

\begin{figure}[!htb]
\centering
\includegraphics[]{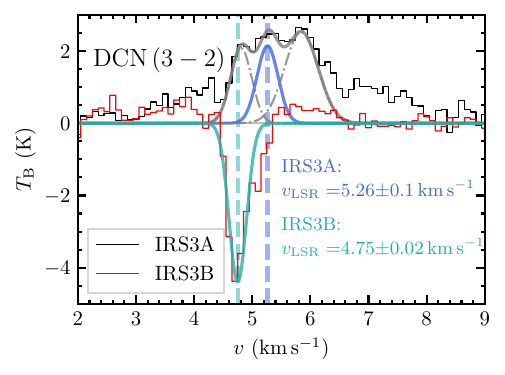}
\caption{Spectrum of DCN ($3-2$) toward the continuum peak position of IRS3A (black) and IRS3B (red). A Gaussian fit to the central velocities are shown in blue and green with source velocities indicated by vertical dashed lines for IRS3A and IRS3B, respectively. In the case of IRS3A, three components were fitted in total to account for the complexity of the spectrum. The central velocity components corresponds to the blue line, while the additional two components are highlighted by the grey dash-dotted lines. The total three-component fit is indicated by the solid grey line.}
\label{fig:DCNspectrum}
\end{figure}

	We use the DCN ($3-2$) transition to estimate the individual source velocities of IRS3A and IRS3B. The spectra of DCN ($3-2$) extracted from the continuum peak position of IRS3A and IRS3B are shown in Fig. \ref{fig:DCNspectrum} in black and red, respectively. The high spectral resolution of the PRODIGE data (0.1\,km\,s$^{-1}$) reveals a complex line profile toward IRS3A with multiple high-velocity components present at both the red- and blueshifted range. From the IRS3A spectrum, we estimate the source velocity to be $5.26\pm0.10$\,km\,s$^{-1}$ from a three-component Gaussian model fit towards the central velocity component. This source velocity is consistent with the results from a position-velocity (PV) diagram of C$^{18}$O ($2-1$) (5.2\,km\,s$^{-1}$, as analyzed in the following) and \citet[][5.4\,km\,s$^{-1}$]{Reynolds2021}. The IRS3A spectrum shows the presence of multiple velocity components and the kinematic properties with a multi-component fit are further analyzed in Sect. \ref{sec:kinematic}.

	The DCN ($3-2$) spectrum toward the continuum peak of IRS3B shows strong absorption which can be explained by the high optical depth of the disk continuum emission and/or foreground DCN with a lower excitation temperature compared to the continuum. High continuum opacities commonly cause absorption features of molecular lines \citep[e.g.,][]{Ohashi2022, Codella2024, Podio2024}. We estimate a source velocity of 4.75$\pm$0.02\,km\,s$^{-1}$ for IRS3B that is consistent with \citet[][4.8\,km\,s$^{-1}$]{Reynolds2021}.
	
	The DCN ($3-2$) emission and absorption toward the continuum peak position of IRS3A and IRS3B, respectively, already reveals the presence of multiple underlying physical components that cause emission or absorption at various velocities. In Sect. \ref{sec:kinematic}, we disentangle the complex line profiles by fitting multiple Gaussian velocity components to DCN ($3-2$) as well as HC$_{3}$N ($24-23$) and CH$_{3}$OH ($4_{2,3}-3_{1,2}E$) to derive the kinematic structure in the entire region with a focus on the gas bridge.

	The mass of a (proto)star is the most important property that will heavily influence its evolution, however, it is not straightforward to measure the protostellar mass of the central objects. \citet{Reynolds2021} estimated the mass of the IRS3B system by obtaining a PV diagram of C$^{17}$O ($3-2$) emission: For the IRS3B-ab binary a mass of 1.15\,$M_\odot$ and for IRS3B-c an upper limit of 0.2\,$M_\odot$ was derived. These authors were not able to reliably fit the PV diagram of IRS3A and estimated a value of 1.4\,$M_\odot$ by a visual inspection. In addition, these authors estimated the IRS3A mass to be $1.5\pm0.1$\,$M_\odot$ by radiative transfer modeling the SO$_{2}$ line emission.
	
	Here, we use the C$^{18}$O ($2-1$) transition of the PRODIGE data to estimate the protostellar mass of IRS3A. For IRS3B, the contribution of the extended envelope and the gas bridge is too high in C$^{18}$O emission in order to reliably estimate the kinematic mass. Young embedded disks around Class 0/I protostars are warm enough to prevent CO freeze-out compared to isolated Class II disks \citep{vantHoff2020}. Radiative transfer models by \citet{Tobin2020} show that the average disk temperature is $\approx$60\,K and $\approx$70\,K for protostellar luminosities of 1\,$L_\odot$ and 10\,$L_\odot$, respectively. Thus C$^{18}$O is expected to be in the gas phase in the entire disk of the embedded Class 0/I IRS3A system with $L=9.2$\,$L_\odot$ \citep{Tobin2016}.

\begin{figure}[!htb]
\centering
\includegraphics[width=0.4\textwidth]{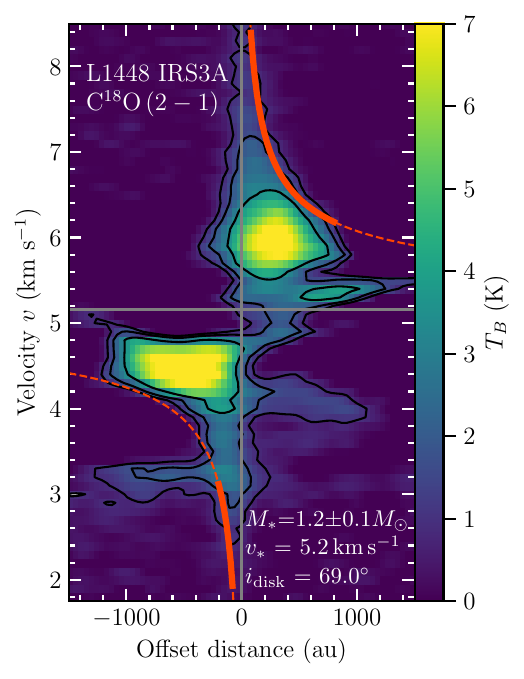}
\caption{Position-Velocity diagram of C$^{18}$O ($2-1$) toward IRS3A extracted along the orientation of the disk with a width of 1$''$. Black contours highlight levels at 4, 8, and 16$\times \sigma_\mathrm{line}$ ($\sigma_\mathrm{line} = 0.22$\,K, Table \ref{tab:obs}). The best-fit derived by \texttt{KeplerFit} is shown as a red dashed line. The ranges in radius included in the fit are indicated by the solid red line.}
\label{fig:pvdiagram}
\end{figure}

	We use the \texttt{pvextractor} package \citep{Ginsburg2015, Ginsburg2016} to extract a PV diagram of C$^{18}$O ($2-1$) along the orientation of the disk \citep[133$\pm$1$^\circ$,][]{Reynolds2021} with a width of 1$''$. The PV diagram is shown in color in Fig. \ref{fig:pvdiagram} whereas 4, 8, and 16$\sigma_\mathrm{line}$ levels are indicated by black contours ($\sigma_\mathrm{line} = 0.22$\,K). In order to estimate the mass, we use the \texttt{KeplerFit} python package \citep{Bosco2019, Bosco2023} to fit the observed PV diagram with models assuming Keplerian disk rotation. A good estimate of the mass, with uncertainties of the order of a few 10\%, can be obtained by fitting the edge of the emission \citep{Seifried2016,Bosco2019,Ahmadi2023}. In our case, we fit the 4$\sigma_\mathrm{line}$ contour level of the emission in the PV diagram. It should be noted though that marginally resolved data can cause an overestimate of the mass due to a broadening along the velocity axis \citep[e.g.,][]{Maret2020}. \citet{Ohashi2023} compare the mass of R CrA IRS7B-a derived with the PV diagram method using the edge method as well as the ridge method and find differences of $\approx$1\,$M_\odot$.
	
	The best-fit Keplerian disk model is indicated by a red dashed line in Fig. \ref{fig:pvdiagram}. We only include ranges in radius that are highlighted by the red solid line that include bright line emission. We mostly focus on the redshifted part of the emission since the blueshifted part is partially suffering from missing flux. The final mass is corrected for the inclination of the disk \citep[$69^\circ$,][]{Reynolds2021}. We assume an uncertainty of $5^\circ$ for the disk inclination. We obtain a mass of 1.2$\pm$0.1\,$M_\odot$, where the uncertainty includes the errors from the fit as well as uncertainty of the inclination. This value is in agreement, within the uncertainties, with the mass reported by \citet[][1.4\,$M_\odot$]{Reynolds2021}. The source velocity from the best-fit Keplerian model is estimated to be 5.2\,km\,s$^{-1}$ which is in agreement with the estimate using the DCN ($3-2$) transition (Fig. \ref{fig:DCNspectrum}).

\subsection{Kinematic structure of the gas bridge in L1448N}\label{sec:kinematic}

\begin{figure*}[!htb]
\centering
\includegraphics[width=0.9\textwidth]{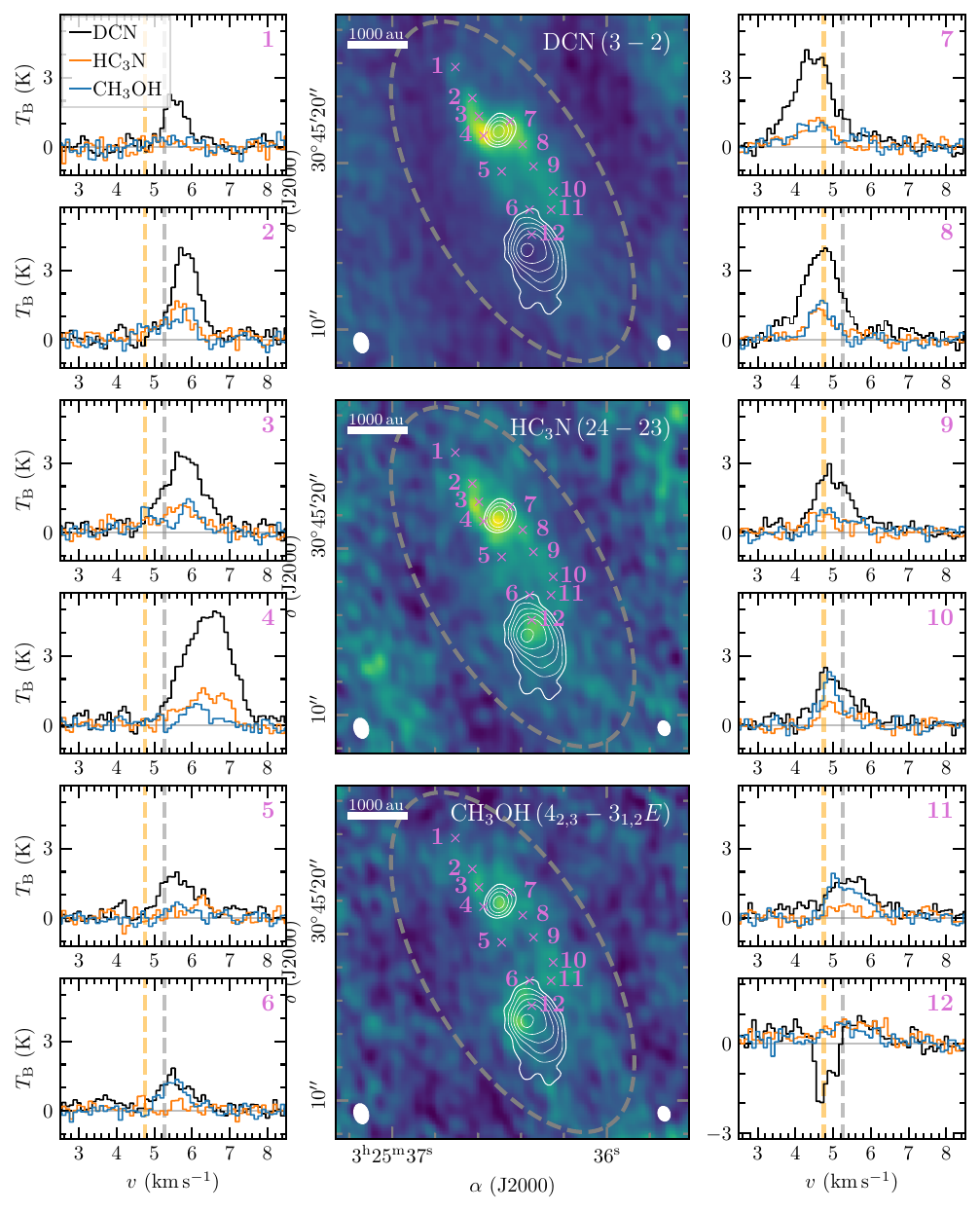}
\caption{Overview of the gas bridge. In the central panels, a zoom in of the integrated intensity maps (Fig. \ref{fig:moment0}) is shown in color (top: DCN $3-2$, middle: HC$_{3}$N $24-23$ and bottom: CH$_{3}$OH $4_{2,3}-3_{1,2}E$). The synthesized beam of the line and continuum data is shown in the bottom left and right, respectively. A scale bar is presented in the top left corner. The grey dashed ellipse shows the area covered by the gas bridge in which the kinematic analysis is conducted. In pink positions are marked and labeled for which spectra are shown in the left and right panels (black: DCN $3-2$, orange: HC$_{3}$N $24-23$ and blue: CH$_{3}$OH $4_{2,3}-3_{1,2}E$). The grey and orange dashed vertical lines indicate the source velocity of IRS3A and IRS3B, respectively.}
\label{fig:bridgeoverview}
\end{figure*}

	The line integrated intensity maps (Fig. \ref{fig:moment0}) reveal an extended gas bridge in L1448N surrounding and embedding the IRS3A and IRS3B systems. The outflow orientations of the systems are known (Sect. \ref{sec:outflow}), we therefore are able to study the gas bridge and its kinematic properties. This gas bridge could be associated with the outflow or there could be a different origin, for example, infalling material. Here, we analyze the kinematic properties the gas bridge using the DCN ($3-2$), HC$_{3}$N ($24-23$), and CH$_{3}$OH ($4_{2,3}-3_{1,2}E$) transitions (bottom row in Fig. \ref{fig:moment0}), as these molecules trace the bridge without contamination from the outflows and the emission is compact enough to not suffer from missing flux. A zoom in on the integrated intensity map of these three molecules and example spectra of these three transitions at selected positions along the gas bridge is presented in Fig. \ref{fig:bridgeoverview}.
	
	The high spectral resolution of the PRODIGE data (0.1\,km\,s$^{-1}$) spectrally resolves the line profiles well and reveals the presence of multiple velocity components as well as clear velocity gradients along the gas bridge. Thus, we first fit up to three Gaussian velocity components to the DCN, HC$_{3}$N, and CH$_{3}$OH line data (Sect. \ref{sec:Gaussianfit}) and then apply a clustering algorithm to disentangle the velocity components (Sect. \ref{sec:clustering}).
	
\subsubsection{Gaussian decomposition of the spectral line profiles}\label{sec:Gaussianfit}

	The kinematic properties, with a focus on the peak velocity of the different velocity components, are analyzed using the \texttt{pyspeckit} python package and its \texttt{specfit} function \citep{Ginsburg2011, Ginsburg2022}. We fit up to one, two, and three Gaussian velocity components ($n_\mathrm{Gauss,max}$) to the HC$_{3}$N, CH$_{3}$OH, and DCN spectra, respectively. Each Gaussian component is defined by three parameters (peak amplitude $I_\mathrm{peak}$, velocity $\varv$, and standard deviation $\sigma_\mathrm{Gauss}$). The corresponding FWHM (full width at half maximum) line width corresponds to $\Delta \varv = 2\sqrt{2\mathrm{ln}(2)} \times \sigma_\mathrm{Gauss}$.

	First, we create a noise map for each transition by estimating the noise in each pixel in line-free channel ranges. Then, the spectrum in each pixel is evaluated and if the S/N is higher than 5 within a velocity range of 3\,km\,s$^{-1}$ and 7\,km\,s$^{-1}$, the spectrum is fed into \texttt{pyspeckit}. We only analyze the area that is covered by the gas bridge, as highlighted by the dashed grey ellipse in Fig. \ref{fig:bridgeoverview}. A single Gaussian component (peak amplitude $I_\mathrm{peak}$, velocity $\varv$, and standard deviation $\sigma_\mathrm{Gauss}$) is then fitted to the observed spectrum using \texttt{specfit} function with \texttt{use\_lmfit $=$ True}. In the case of the HC$_{3}$N ($24-23$) transition, no significant residuals were present in the spectra with one fitted Gaussian velocity component.
	
	In a next step, for DCN and CH$_{3}$OH, a second Gaussian velocity component is fitted to the data. This procedure is repeated once more for DCN, given that the velocity profile of DCN ($3-2$) shows areas that are clearly dominated by three velocity components (Fig. \ref{fig:bridgeoverview}). In Fig. \ref{fig:examplefit} example spectra as well as the best fit from the Gaussian decomposition are presented along some positions along the gas bridge (Fig. \ref{fig:bridgeoverview}). 
	
	To avoid over-fitting as well as discard unreliable fits, we apply a quality assessment of the results. We discard for all three transitions all velocity components with $\sigma_\mathrm{Gauss} < 0.07$\,km\,s$^{-1}$ (corresponding to $\Delta \varv = 0.16$\,km\,s$^{-1}$) which are associated to unreliable narrow fits in noisy spectra. For DCN and CH$_{3}$OH, we compute the Akaike information criterion (AIC) for the one-, two- and (in the case of DCN) three-component fit (AIC$_1$, AIC$_2$, and AIC$_3$, respectively) in order to select the best model fit. The AIC considers the least $\chi^2$ statistics as well as the number of free parameters in each model \citep[see Appendix B in][]{ValdiviaMena2022}. The best model is the one with the lowest value of AIC, AIC$_\mathrm{min}$, but only if it improves the AIC value by a difference of at least $\Delta$AIC = 5. The median uncertainty of the FWHM line width is 0.09, 0.07, and 0.11\,km\,s$^{-1}$ for DCN, HC$_{3}$N, and CH$_{3}$OH results, respectively.
	
	Given the complexity of the velocity components and the spatial overlap, it is not straightforward to assign the velocity components to their underlying physical structure. Thus, in the next section, we use a clustering algorithm to disentangle the velocity components.

\subsubsection{Clustering of the velocity components}\label{sec:clustering}

\begin{figure*}[!htb]
\centering
\includegraphics[]{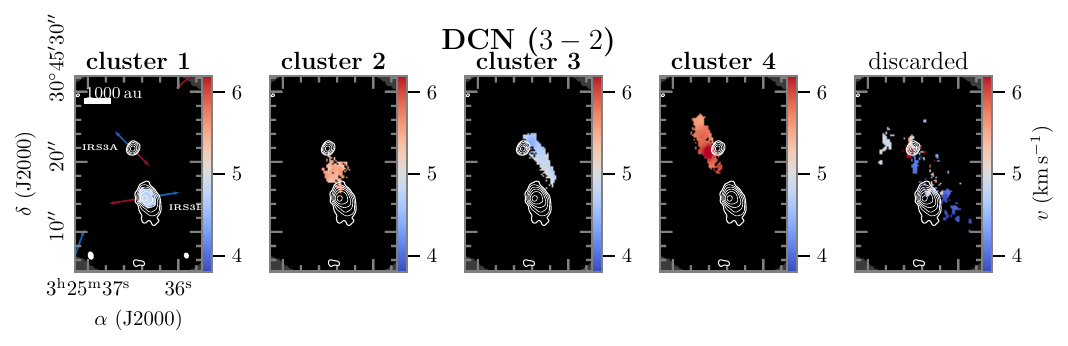}\\
\caption{Velocity clusters derived with \texttt{DBSCAN} for DCN ($3-2$). The velocity of a cluster is shown in each panel in color, while the discarded data points are shown in the last panel. The 1.4\,mm continuum data is presented in white contours and levels are 5, 10, 20, 40, 80, 160$\times \sigma_\mathrm{cont}$ ($\sigma_\mathrm{cont}$=0.94\,mJy\,beam$^{-1}$). In the first panel, the outflow directions are indicated and sources are marked. A scale bar of 1000\,au is shown in the top left. The synthesized beam of the line and continuum data is shown in the bottom left and right, respectively. The velocity clusters for HC$_{3}$N and CH$_{3}$OH are presented in Fig. \ref{fig:clusters_app}. The corresponding amplitude $I_\mathrm{peak}$ and FWHM line width $\Delta \varv$ of all three transitions are presented in Fig. \ref{fig:clusters_app_ampli} and Fig. \ref{fig:clusters_app_width}, respectively.}
\label{fig:clusters}
\end{figure*}

	We use the density-based spatial clustering algorithm \texttt{DBSCAN} (Density-Based Spatial Clustering of Applications with Noise) of the \texttt{scikit-learn} python package \citep{Grisel2021} to cluster the velocity components for each transition (DCN, HC$_{3}$N, CH$_{3}$OH). We aim to recover the kinematic signature of the gas bridge.
	
	Each cluster derived by \texttt{DBSCAN} consists of central core points as well as nearby neighborhood points. The clustering with \texttt{DBSCAN} is set by two parameters: The maximum distance between two data points for one to be considered as in the neighborhood of the other, \texttt{eps}, and the minimum number of data points in a cluster, \texttt{minsamples}. With \texttt{DBSCAN}, some data points may not be associated with any cluster and we refer to those as discarded data points. The input parameters for \texttt{DBSCAN} are summarized in Table \ref{tab:dbscan} for each of the three transitions. 
	
	The input for \texttt{DBSCAN} is a list of data points that each consists of the 2D position and velocity ($x$,$y$,$\varv$). The peak intensity and line width are not given as an input, but we will also analyze the spatial distribution of these parameters of all velocity clusters to check that \texttt{DBSCAN} provides reasonable results. To further minimize the effect of outliers, the data set is scaled using the \texttt{RobustScaler} function of \texttt{scikit-learn}. This function removes the median and scales the data according to the quantile range between the first and third quantile. Then we run \texttt{DBSCAN} for each molecule on this scaled data set. The output of \texttt{DBSCAN} is the number of clusters, as well as a list of data points ($x$,$y$,$\varv$) for each cluster. Data points that cannot be associated with any cluster according to the defined parameters (\texttt{eps} and \texttt{minsamples}), are discarded.
	
	The results from \texttt{DBSCAN} are summarized in Table \ref{tab:dbscan} and Fig. \ref{fig:clusters} shows the velocity map of each cluster for DCN, while the results for HC$_{3}$N and CH$_{3}$OH are presented in Fig. \ref{fig:clusters_app}. For DCN, there is a minor fraction of pixels after the clustering that contain two velocity components, in these cases, the average values of the fit parameters (velocity, peak amplitude, and line width) are calculated. The peak amplitude and line width maps of the clusters of all three transitions are presented in Fig. \ref{fig:clusters_app_ampli} and Fig. \ref{fig:clusters_app_width}, respectively.
	
	For the DCN $3-2$ transition (Fig. \ref{fig:clusters}) we find in total four velocity clusters with \texttt{DBSCAN}. Cluster 1 traces the emission around the IRS3B disk at the source velocity of $\approx$4.7\,km\,s$^{-1}$. Clusters 2, 3, and 4 reveal the extended gas bridge and the presence of a red- (cluster 2 and 4) and blueshifted (cluster 3) part (in the east and west, respectively). At both sides of the bridge, the velocities increase toward IRS3A with respect to the source velocity of $\approx$5.2\,km\,s$^{-1}$. In Sect. \ref{sec:velocitygradient} we will further discuss the velocity gradients of the bridge. Most notably, the envelope/disk rotation of IRS3A \citep[as revealed by C$^{17}$O $3-2$,][]{Reynolds2021} shows the same velocity pattern as the gas bridge (redshifted east of IRS3A and blueshifted west of IRS3A). With the angular resolution of the PRODIGE data (300\,au), we cannot resolve the IRS3A disk with a size of $\approx$100\,au. Thus high angular molecular line observations are required to better study how the bridge structure connects to the IRS3A disk.
	
	In total four clusters are found with \texttt{DBSCAN} for HC$_{3}$N $24-23$ (top panel in Fig. \ref{fig:clusters_app}). Cluster 1 and 4 can be associated with the blue- and redshifted part of the gas bridge, while clusters 2 and 3 trace the red and blueshifted emission of the disk/envelope rotation of IRS3A. The HC$_{3}$N emission is much less extended compared to DCN ($3-2$). Given the relatively high upper energy level of 131\,K for HC$_{3}$N (Table \ref{tab:obs}), the emission might trace warmer locations along the gas bridge as well as in the envelope/disk region.
	
	For the CH$_{3}$OH $4_{2,3}-3_{1,2}E$ transition, we find four velocity clusters with \texttt{DBSCAN}. Cluster 1 and 2 can be associated with envelope rotation of IRS3B. The velocity pattern roughly matches the C$^{17}$O ($3-2$) envelope rotation found by \citet{Reynolds2021}. We only detect CH$_{3}$OH emission toward the outer edges of the disk. The non-detection towards the bright continuum disk location is most likely due to the high optical depth of the continuum blocking the line emission \citep[e.g.,][]{DeSimone2020}. Cluster 3 and 4 can be associated with the blue- and redshifted part of the gas bridge, respectively.
	
	In the right panels in Figs. \ref{fig:clusters} and \ref{fig:clusters_app} we present the velocity maps of the discarded data. Given the small separation in both position and velocity, it is, with the spatial resolution of the PRODIGE data, challenging to disentangle the velocity components close to the IRS3A disk/envelope system as well as the spatial intersection of the red and blueshifted part of the gas bridge.
	
	The amplitude $I_\mathrm{peak}$ and FWHM line width $\Delta \varv$ maps of all clusters derived with the three species are presented in Figs. \ref{fig:clusters_app_ampli} and \ref{fig:clusters_app_width}, respectively. Given the smooth maps of the peak intensity and line width, both of which were no input parameters for the clustering, confirms that the clustering of the velocity data points with \texttt{DBSCAN} provides reliable results. 
	
	In the DCN ($3-2$) transition, the line widths along the gas bridge are narrow and on the order of $\approx$1\,km\,s$^{-1}$ or below, but there is an increase toward the location of the unresolved IRS3A disk. Given that the line width of CH$_{3}$OH along the bridge is as narrow as DCN and HC$_{3}$N, the emission is most likely not caused by shocks. In order to disentangle the region close to the IRS3A protostar and its disk, line observations at higher angular resolution are necessary.

\section{Discussion}\label{sec:discussion}

	We have revealed with the PRODIGE molecular line data of IRS3A and IRS3B located in L1448N an extended gas bridge between these two protostellar systems. We have disentangled the velocity components in the spectra to extract the kinematic signature of the gas bridge in the molecular tracers DCN ($3-2$), HC$_{3}$N ($24-23$), and CH$_{3}$OH ($4_{2,3}-3_{1,2}E$) in Sect. \ref{sec:kinematic}. We find that the gas bridge consists of a redshifted side toward the east and a blueshifted side west of the IRS3A and IRS3B systems. In the following, we discuss the properties of the gas bridge, i.e. velocity gradients and infalling motions, the impact of infall, and possible origins of the gas bridge.

\subsection{Infall motions along the gas bridge}\label{sec:infall}

\subsubsection{Velocity gradients}\label{sec:velocitygradient}

\begin{figure*}[!htb]
\centering
\includegraphics[width=0.8\textwidth]{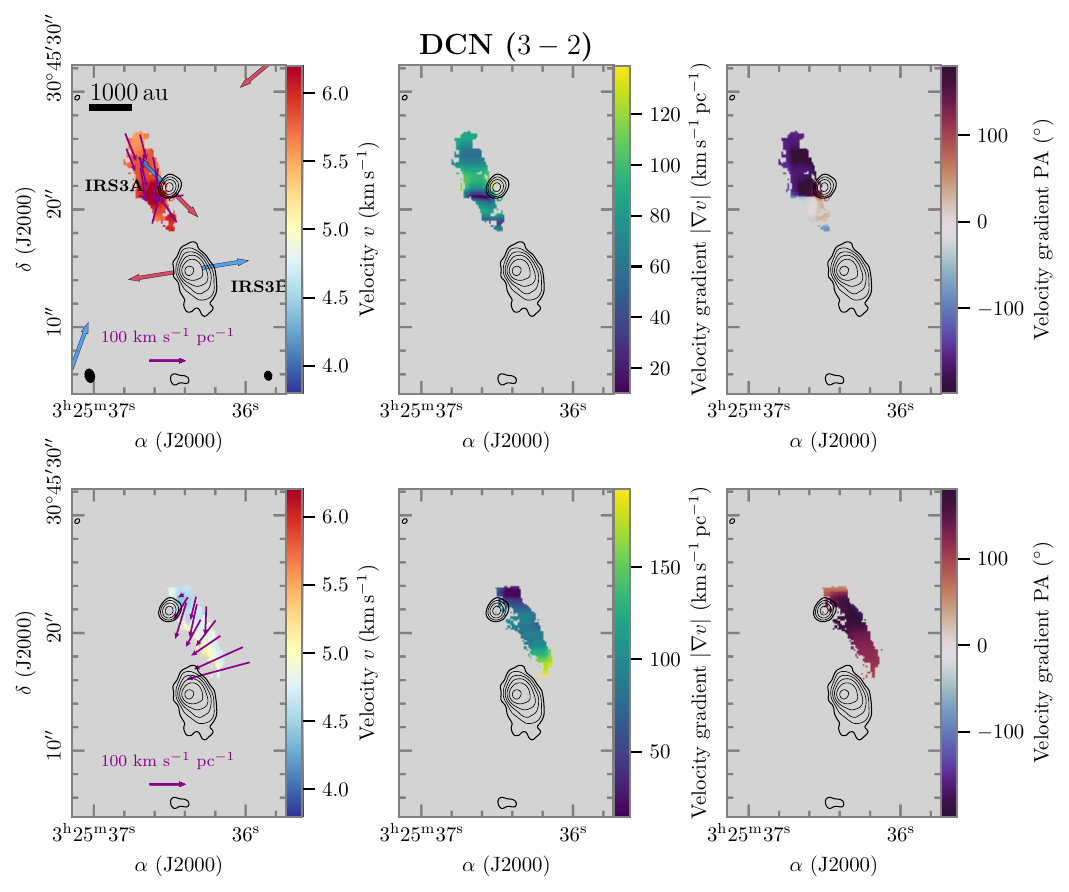}
\caption{Velocity gradient maps of the red (top) and blueshifted (bottom) part of the gas bridge. In each row the velocity (left), velocity gradient (center), and velocity gradient PA (right) is shown in color. The black contours are the 1.4\,mm continuum with levels at 5, 10, 20, 40, 80, 160$\times \sigma_\mathrm{cont}$ ($\sigma_\mathrm{cont}$=0.94\,mJy\,beam$^{-1}$). The purple arrows in the left panels indicate the direction and strength of the velocity gradient, where the arrow tip points from low to high velocities. The length of the purple arrow in the bottom left marks a velocity gradient of 100\,km\,s$^{-1}$\,pc$^{-1}$. In the top left panel, the synthesized beam of the NOEMA line and continuum data is shown in the bottom left and right, respectively, a scale bar is shown in the top left corner, and bipolar outflow orientations are highlighted by red and blue arrows (Sect. \ref{sec:outflow}).}
\label{fig:velocity_gradient}
\end{figure*}

	To study velocity gradients along both sides of the gas bridge, we use the velocity gradient code used by \citet{Pineda2010} and available in the \verb+velocity_tools+\footnote{\url{https://github.com/jpinedaf/velocity_tools}} python package. The velocity maps of the red and blueshifted parts of the gas bridge already show velocity gradients along its major axis (Figs. \ref{fig:clusters} and \ref{fig:clusters_app}). Whereas the difference in velocity is only on the order of a few 0.1\,km\,s$^{-1}$ - highlighting the necessity for a high spectral resolution - this difference occurs over small spatial scales, on the order of a few 1000\,au. We compute the velocity gradient using the velocity maps of both sides of the gas bridge derived with \texttt{DBSCAN} (Sect. \ref{sec:clustering}). The results for DCN ($3-2$) are shown in Fig. \ref{fig:velocity_gradient}, while the results for HC$_{3}$N ($24-23$) and CH$_{3}$OH ($4_{2,3}-3_{1,2}E$) are shown in Figs. \ref{fig:velocity_gradient_app1} and \ref{fig:velocity_gradient_app2}, respectively. The velocity gradient results for the red and blueshifted part of the gas bridge are shown in the top and bottom panels in Fig. \ref{fig:velocity_gradient}, respectively. The left panel is the velocity map overlaid with purple arrows that highlight the direction and strength of the velocity gradient, the middle panel shows the velocity gradient map, and the right panel is a map of the PA of the velocity gradient.
	
	The velocity gradient of the gas bridge is on the order of a few 10\,km\,s$^{-1}$\,pc$^{-1}$ to $\approx$120\,km\,s$^{-1}$\,pc$^{-1}$ for DCN ($3-2$). For both parts of the gas bridge, the velocities become more extreme toward IRS3A, i.e., more redshifted and blueshifted. This suggests infalling motions from both parts of the gas bridge toward IRS3A. In the next section, we apply a streamline model to both sides of the gas bridge to check if gravitational infall caused by the IRS3A protostar can explain the observed velocities along the gas bridge. The southern part of the redshifted gas bridge hints at velocity gradients toward IRS3B as well which can also be clearly seen in the change in the PA of the gradient near IRS3B (top right panel in Fig. \ref{fig:velocity_gradient}). More sensitive and higher angular resolution observations are required in order to investigate a potential connection of the gas bridge toward the IRS3A and IRS3B disks.
	
	Even though the velocity maps of HC$_{3}$N and CH$_{3}$OH are less extended compared to DCN ($3-2$), the redshifted side of the gas bridge shows a similar morphology in velocity as DCN (Figs. \ref{fig:velocity_gradient_app1} and \ref{fig:velocity_gradient_app2}). The bottom part of the blueshifted side of the gas bridge in CH$_{3}$OH shows that there is also an additional east-west gradient in the direction of IRS3B. While the gas bridge shows increasing velocity gradients toward IRS3A, there might be infall toward IRS3B as well. 
	
	Notably, the rotation of the IRS3A disk/envelope system \citep{Reynolds2021} occurs in a similar direction, i.e., west (blueshifted) to east (redshifted). This is a strong indication that the gas bridge is a remnant structure of the common elongated filament out of which the IRS3A and IRS3B systems have formed. We further discuss the origin of the gas bridge in Sect. \ref{sec:originridge}.
	
	The overall, i.e. global, velocity gradient for each side and molecular tracer of the gas bridge is summarized in Table \ref{tab:dbscan}. We find global velocity gradients in the range of 50$-$120\,km\,s$^{-1}$\,pc$^{-1}$ and 70$-$130\,km\,s$^{-1}$\,pc$^{-1}$ for the redshifted and blueshifted side of the gas bridge, respectively. These values are consistent with velocity gradients found by \citet[][70\,km\,s$^{-1}$\,pc$^{-1}$ and 120\,km\,s$^{-1}$\,pc$^{-1}$ for IRS3A and IRS3B, respectively]{Gupta2022} that studied the velocity gradients using C$^{18}$O ($2-1$). One explanation for the differences in both sides might be due to the unknown inclination of the gas bridge. In the scenario of the infalling gas bridge, a high velocity gradient could cause strong accretion events in the IRS3A disk and can explain why IRS3A is much brighter in the IR compared to IRS3B \citep{OLinger2006}. In contrast to this gas bridge around IRS3A and IRS3B, the gas bridge toward IRAS\,$16293-2422$ is quiescent with no strong velocity gradients seen in C$^{17}$O ($3-2$) \citep{vanderWiel2019}.

\subsubsection{Streamline model}\label{sec:streamlinemodel}

\begin{figure*}[!htb]
\centering
\includegraphics[width=0.4\textwidth]{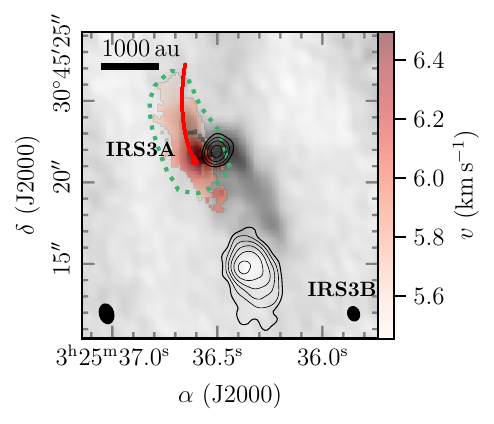}
\includegraphics[width=0.4\textwidth]{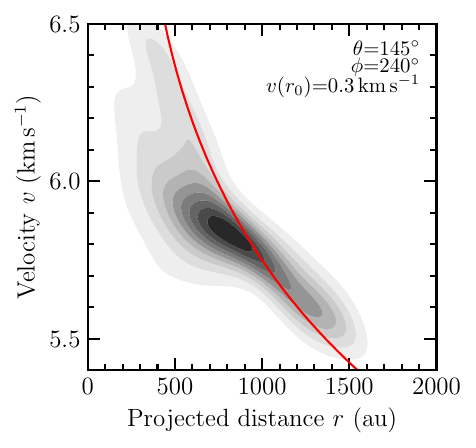}\\
\includegraphics[width=0.4\textwidth]{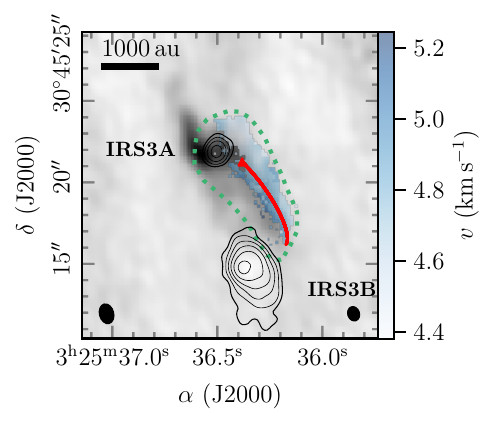}
\includegraphics[width=0.4\textwidth]{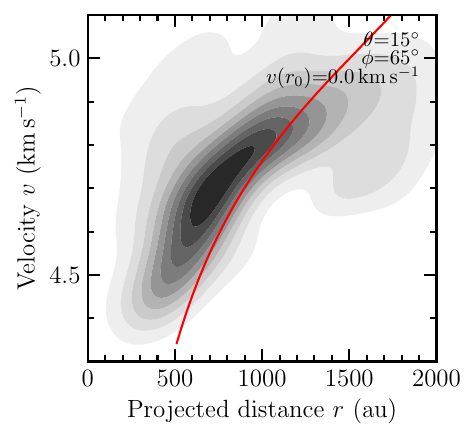}
\caption{Streamline model of the red- (top) and blueshifted (bottom) side of the gas bridge. The left panel shows the integrated intensity of DCN ($3-2$) in greyscale and the velocity map in color. The black contours are the 1.4\,mm continuum with levels at 5, 10, 20, 40, 80, 160$\times \sigma_\mathrm{cont}$ ($\sigma_\mathrm{cont}$=0.94\,mJy\,beam$^{-1}$). The synthesized beam of the NOEMA line and continuum data is shown in the bottom left and right, respectively. A scale bar is shown in the top left. The right panel shows the kernel density estimate of the observed velocity profile in greyscale as a function of projected distance and the streamline model in red. The projected trajectory of the streamline model is highlighted in red in the left panels as well. The position where the streamline model reaches the centrifugal radius is marked by the red triangle.}
\label{fig:streamline_model}
\end{figure*}

	To investigate if the observed velocity gradients along both sides of the gas bridge can be explained by gravitational infall due to the central IRS3A protostar, we use the streamline model by \citet{Pineda2020}. The model uses an analytic solution of an free-falling accretion flow of a rotating parcel toward a central gravitating object assuming conservation of angular momentum \citep{Mendoza2009}. We apply the model to the observed DCN ($3-2$) velocities of both the red- and blueshifted side (cluster 4 and 3, respectively) of the gas bridge derived in Sect. \ref{sec:clustering}.
	
	The streamline model computes the position and velocity of a test mass along the streamline. The coordinate system is set such that the z-axis points along the angular momentum vector of the disk and the x- and y-axis correspond to the disk plane. For both streamline models, we use an inclination $i$ of 21$^\circ$ \citep{Reynolds2021} and position angle of 225$^\circ$ (Sect. \ref{sec:outflow}) to set the orientation of the angular momentum vector of the disk. The protostellar mass is 1.2\,$M_\odot$ and the source velocity is set to 5.26\,km\,s$^{-1}$, as derived in Sect. \ref{sec:PV}.
	
	The accretion flow is described by its initial position ($r_0$, $\theta_0$, $\phi_0$) and initial radial velocity $\varv_{r0}$. Due to our limited FoV set by the primary beam, we set $r_0 = 2500$\,au for both sides of the gas bridge that matches the extent of the observed gas bridge structure. The initial rotation of the parcel is described by $\Omega_0$, which we estimate by the difference of peak velocity of the gas bridge with respect to the source velocity that is $\approx0.5$\,km\,s$^{-1}$ and the length of the streamer, $\Omega_0 \approx \frac{\Delta \varv}{r_0} \approx 1.3\times10^{-12}$\,s$^{-1}$. The corresponding centrifugal radius is $r_\mathrm{centr} = 1200$\,au. The streamline model computes the position and velocity from $r_0$ up until $r_\mathrm{centr}$.
	
	We explore the parameter space of the remaining three parameters - $\theta_0$, $\phi_0$, and $\varv_{r0}$ - and search manually by eye for streamline models that reproduce both the projected trajectory as well as velocity profile. The left panels in Fig. \ref{fig:streamline_model} shows the integrated intensity of DCN ($3-2$) in greyscale and overlaid the red- and blueshifted velocity side of the gas bridge in color in the top and bottom, respectively. The right panels shows the corresponding kernel density estimate of the observed velocity profile in greyscale. In the left panels the projected trajectory and in the right panels the velocity profile of the best match streamline model is shown by the red line. For the redshifted side of the gas bridge we obtain a good match for $\theta_0 = 145^\circ$, $\phi_0 = 240^\circ$, and $\varv_{r0}=0.3$\,km\,s$^{-1}$ and for the blueshifted side of the gas bridge we obtain a good match for $\theta_0 = 15^\circ$, $\phi_0 = 65^\circ$, and $\varv_{r0}=0.0$\,km\,s$^{-1}$.
	
	For both sides of the gas bridge we find solutions of the streamline model that match the observed trajectory as well as velocity profile. Thus, the observed velocity gradients can be caused by the gravitational pull of the central IRS3A protostar. Limitations of the streamline model include the assumption of conservation of angular momentum and that no interaction with the surrounding envelope material is considered. The impact of the infalling material is further discussed in Sect. \ref{sec:impactdisk}.

\subsection{Impact of the gas bridge onto the IRS3A and IRS3B disks}\label{sec:impactdisk}

	We find that the gas bridge connects to the rotation disk/envelope of IRS3A with infalling motions. There is evidence that the bridge also partially connects to the IRS3B disks, however, higher angular resolution data are required to disentangle the kinematically complex region. The PRODIGE data at 1$''$ ($\sim$300\,au) resolution is not sufficient to spatially resolve the IRS3A disk, thus higher angular resolution observations are necessary to resolve the impact of the gas bridge onto the disk. Compact emission of shock-tracing species such as SO and OCS are bright toward the position of the IRS3A disk (Fig. \ref{fig:moment0}). Asymmetric SO emission has already been attributed to a streamer impact toward Per-emb-50 within the PRODIGE data \citep{ValdiviaMena2022}.
	
	Ring-like and spiral arm structures are evident in the disks of IRS3A and IRS3B, respectively \citep{Reynolds2021, Reynolds2024}. Gravitational instabilities could have been triggered through the infalling material of the gas bridge as suggested by simulations \citep[e.g.,][]{Kuffmeier2018}. Comparing our large scale observations with the high angular resolution ALMA observations by \citet{Reynolds2021} hint at a connection of the gas bridge with the IRS3A disk in H$^{13}$CO$^{+}$ and SO$_{2}$ emission (their Figs. 8 and 10, respectively). These two transitions shows the same red- and blueshifted morphology of the gas bridge in the east and west, respectively, as seen in DCN, HC$_{3}$N, and CH$_{3}$OH. The integrated intensity of CS $5-4$ from the ALMA PEACHES survey also shows the redshifted side of the gas bridge and is interpreted as an infalling envelope \citep{ArturdelaVillarmois2023}, while in the maps of IRS3B the CS emission follows the bipolar outflow. Not only high spatial resolution observations in the mm regime of shock tracing molecules, but sensitive MIR observations of hydrogen recombination lines with the James Webb Space Telescope (JWST), could also reveal the presence of accretions shocks in the IRS3A disk \citep{Rigliaco2015}.

\subsection{Possible origins of the gas bridge}\label{sec:originridge}

\begin{figure}[!htb]
\centering
\includegraphics[width=0.4\textwidth]{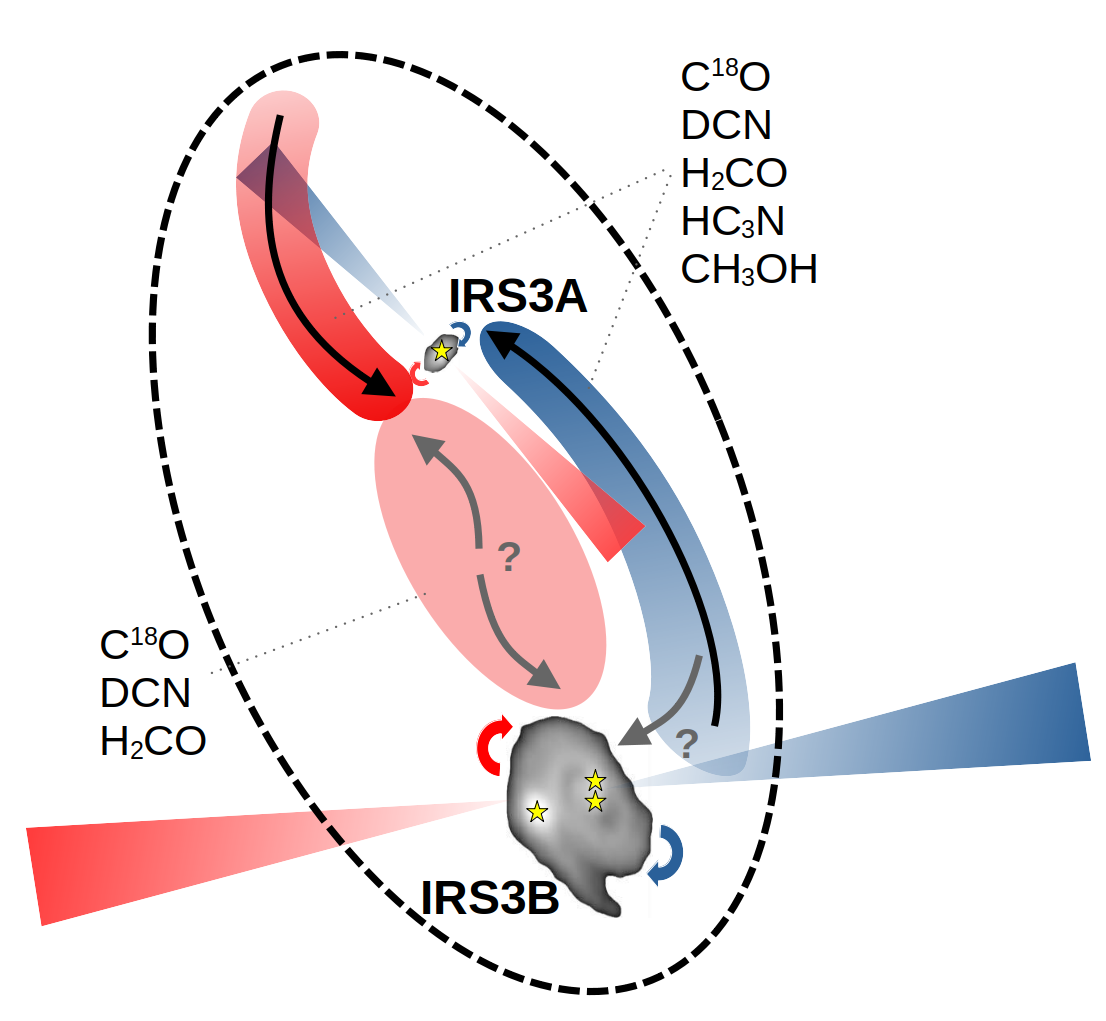}
\caption{Sketch of the kinematic components of the gas bridge as well as bipolar outflows surrounding IRS3A and IRS3B protostellar systems in L1448N. The greyscale images shows the 1.3\,mm emission with ALMA are taken from the VANDAM survey \citep{Tobin2018}. Individual protostars are highlighted by the yellow stars.}
\label{fig:sketch}
\end{figure}

	A kinematic analysis of the PRODIGE molecular line data reveals that the gas bridge is infalling towards IRS3A from both sides of the gas bridge (Sect. \ref{sec:infall}). A sketch of infall and outflow motions is presented in Fig. \ref{fig:sketch}. In the following, we discuss potential explanations of how the bridge structure formed.
	
\subsubsection{Molecular outflow of IRS3A}

	The orientation of the gas bridge is parallel to the orientation of the bipolar outflow from IRS3A. However, we find that the spatial morphology and kinematic properties of CO tracing the outflow are different from the gas bridge traced in DCN, HC$_{3}$N, and CH$_{3}$OH. The line widths of these tracers are narrow ($<$2\,km\,s$^{-1}$, Fig. \ref{fig:clusters_app_width}) and thus likely not caused by shocks due to the IRS3A outflow. We can thus rule out that the gas bridge a signature of a rotating outflow as it has been suggested for other protostars with rotating outflows \citep[e.g.,][]{Zapata2015, Bjerkeli2016, Tabone2017, Zhang2018, Oya2021, Nazari2024}.

\subsubsection{Remnant filamentary material}

	Another explanation is that the gas bridge is remnant material from the natal envelope structure from which the IRS3A and IRS3B systems have formed. SMA observations (with a primary beam 2.5$\times$ larger compared to NOEMA at 1.4\,mm) of the MASSES survey reveal an extended N-S filament connecting IRS3A and IRS3B in $^{13}$CO ($2-1$), C$^{18}$O ($2-1$), as well as HCO$^{+}$ ($4-3$) emission \citep[][their Fig. 5]{Lee2015}. The IRS3A and IRS3B systems can accrete material from this filament. While \citet{OLinger2006} claim that IRS3A is IR-bright ($S_{12.5\upmu\mathrm{m}} = 0.43$\,Jy compared to $S_{12.5\upmu\mathrm{m}} < 0.12$\,Jy for IRS3B) due to the fact that it has already dispersed the surrounding material, the results by \citet{Lee2015} and our PRODIGE data actually reveal that both IRS3A and IRS3B are still embedded within filamentary material (Fig. \ref{fig:moment0}). The high IR brightness of IRS3A can be explained by the fact that it is accreting from the gas bridge (Sec. \ref{sec:streamlinemodel}) causing a high accretion luminosity. Numerical simulations of low-mass star-forming regions commonly show that protostars in multiple systems are surrounded by filamentary gas structures \citep[e.g.,][]{Kratter2010, Kuffmeier2018, Kuffmeier2019, Lee2019}.
	
	Simulations by \citet{Kuffmeier2019} reveal a bridge structures (with sizes of $10^3-10^4$\,au) connecting protostars that formed as a result of compressive flows in turbulent environments. Protostars accrete material from the bridge feeding the disks. In the simulations, the bridge structures are transient objects that last a few 10\,kyr and are connected to the larger-scale filament. In addition, turbulent fragmentation causes the formation of protostellar companions with separations of $\sim$1000\,au. Comparing these simulations with the observed properties of the IRS3A and IRS3B systems, the close binary system IRS3B-a/b might have formed from disk fragmentation \citep[e.g.,][]{Kratter2010}, whereas the tertiary component may have formed later through turbulent fragmentation \citep[e.g.,][]{Offner2010} of the gas bridge and migrated towards IRS3B-a/b system. The simulated gas bridge by \citet{Kuffmeier2019} could provide material toward multiple protostars and in the case of the L1448N system, both the IRS3A and IRS3B-c protostars could still be heavily accreting from the gas bridge. An analysis of the stability of the IRS3B disk however suggests that the IRS3B-c protostar could have also formed via disk fragmentation \citep{Tobin2016b, Reynolds2021}. Higher angular resolution observations of molecular lines toward IRS3B are necessary to study the impact of the gas bridge onto the disk.
	
\subsubsection{Triggered star formation}

	L1448N is located in the north-west of L1448-mm (Fig. \ref{fig:continuum}) with the blueshifted outflow-lobe of L1448-mm not only oriented along L1448N, but we also detect blueshifted emission knots from CO, SO, H$_{2}$CO, and CH$_{3}$OH in the PRODIGE FoV (Fig. \ref{fig:moment0}). It was already hypothesized by \citet{Barsony1998} that L1448-mm could have triggered the formation of L1448N systems. Given that the redshifted lobe of the IRS3C outflow impacts the L1448 IRS3A and IRS3B systems as well (Fig. \ref{fig:outflow}), the material from the L1448-mm and IRS3C systems could have affected the material around IRS3A and IRS3B from both sides. The impact of the outflows might have shocked and compressed the filament toward IRS3A and IRS3B, while in the surroundings, cold material evident in N$_{2}$H$^{+}$ emission that is spatially offset from the filament connecting IRS3A and IRS3B \citep{Lee2015}. High angular resolution in the MIR with JWST targeting heated and/or shocked material are necessary to follow-up on this. Spitzer and Herschel observations show bright IR emission from atomic and molecular lines such as H$_{2}$, Fe{\sc ii}, S{\sc i}, O{\sc i} originating from the L1448-mm outflow \citep{Neufeld2009, Nisini2013, Nisini2015}. However, the spatial resolution is not sufficient to disentangle the different protostellar systems within L1448N, but it is still a possibility that the L1448-mm outflow could have influenced or even triggered the formation of protostars in L1448N. To further investigate this, a better determination of the L1448-mm outflow timescale with respect to the lifetimes of the protostellar systems in L1448N is necessary.
	
	The protostellar systems in L1448N are surrounded by nearby systems with large-scale outflows. The molecular gas in L1448N shows complex kinematic properties. Extended gas is revealed in a bridge structure surrounding IRS3A and IRS3B is most likely connected to the filamentary structure from which these systems have formed. However, the L1448-mm outflow as well as the outflows of IRS3A could have also influenced the formation of the bridge structure.

\section{Conclusions}\label{sec:conclusions}

	NOEMA 1\,mm observations of the PRODIGE project toward L1448N reveal extended molecular emission in the structure of a gas bridge, surrounding the IRS3A and IRS3B protostellar systems. We analyzed both the spatial morphology as well as the kinematic properties of spectral lines in several molecular tracers, such as DCN, HC$_{3}$N, and CH$_{3}$OH.

	\begin{enumerate}
		\item The PRODIGE data reveal that the IRS3A and IRS3B systems are embedded within extended molecular gas that extends well beyond the FoV of the observed mosaic ($>$6600\,au). An elongated gas bridge with bright emission connects the IRS3A and IRS3B protostellar systems, most evident in C$^{18}$O, H$_{2}$CO, DCN, HC$_{3}$N, and CH$_{3}$OH.
		\item Using a Keplerian disk rotation model applied to the observed PV diagram of C$^{18}$O, we are able to determine the central mass of IRS3A to be 1.2$\pm$0.1\,$M_\odot$. This value is in agreement with radiative transfer modeling results by \citet{Reynolds2021} that estimate the mass to be $1.5\pm0.1$\,$M_\odot$.
		\item The high spectral resolution data with a channel width of 0.1\,km\,s$^{-1}$ reveal complex dynamics including the presence of multiple velocity components along the line of sight. For DCN, HC$_{3}$N, and CH$_{3}$OH, individual velocity components are extracted with a multi-component Gaussian fitting. By applying the \texttt{DBSCAN} clustering algorithm, these velocity components are associated with disk/envelope rotation of the protostellar systems and with a red and blueshifted side of the gas bridge.
		\item Velocity gradients along the gas bridge reveal infalling material toward IRS3A from both the red- and blueshifted sides. Locally, the velocity gradients rise up to 120\,km\,s$^{-1}$\,pc$^{-1}$, while the global average is $\approx$100\,km\,s$^{-1}$\,pc$^{-1}$. We apply a streamline model to the observed velocity profiles of both sides of the gas bridge and confirm that the observations are consistent with gravitational infall toward IRS3A.
		\item Both IRS3A and IRS3B launch bipolar outflows as revealed by CO emission. The IRS3B outflow is also detected in high-velocity CO and SiO bullets. Within the observed FoV there is also a contribution from the nearby L1448-mm and IRS3C outflows (blueshifted and redshifted lobe, respectively). It has been suggested in the past that the L1448-mm outflow could influence the IRS3A and IRS3B systems. We find indeed bright SiO emission, hence shocked locations, caused by L1448-mm toward L1448N.

	\end{enumerate}
	
	Very high angular resolution observations often filter out extended emission. In L1448N harboring the IRS3A and IRS3B system, we have revealed with NOEMA 1.3\,mm observations extended molecular emission surrounding both systems. The kinematic properties of the extended gas bridge are complex, and we find that the most likely explanation for the origin of the structure is that this is the remnant filament material out of which both IRS3A and IRS3B systems have formed. To study the full extent of the gas bridge, a larger area needs to be covered in follow-up observations. At the same time, to reveal the impact zone of the gas bridge onto the IRS3A and IRS3B disks, more sensitive spectral line observations at higher angular resolution are required. Nevertheless, the PRODIGE data highlights the importance of studying the extended, often infalling, material surrounding protostellar systems on scales of a few 100\,au.

\begin{acknowledgements}
	We thank the referee John Tobin for his constructive feedback that substantially improved the paper. The authors thank the IRAM staff at the NOEMA observatory for their support in the observations and data calibration. This work is based on observations carried out under project number L19MB with the IRAM NOEMA Interferometer. IRAM is supported by INSU/CNRS (France), MPG (Germany) and IGN (Spain). DSC is supported by an NSF Astronomy and Astrophysics Postdoctoral Fellowship under award AST-2102405. D. S. and Th. H. acknowledge support from the European Research Council under the Horizon 2020 Framework Program via the ERC Advanced Grant Origins 832428 (PI: Th. Henning). I.J-.S acknowledges funding from grants PID2022-136814NB-I00 funded by MICIU/AEI/ 10.13039/501100011033 and by "ERDF/EU".
\end{acknowledgements}

\bibliographystyle{aa}
\bibliography{bibliography}

\begin{appendix}

\section{Spectra of L1448N}\label{sec:spectra}

\begin{figure*}[!htb]
\centering
\includegraphics[]{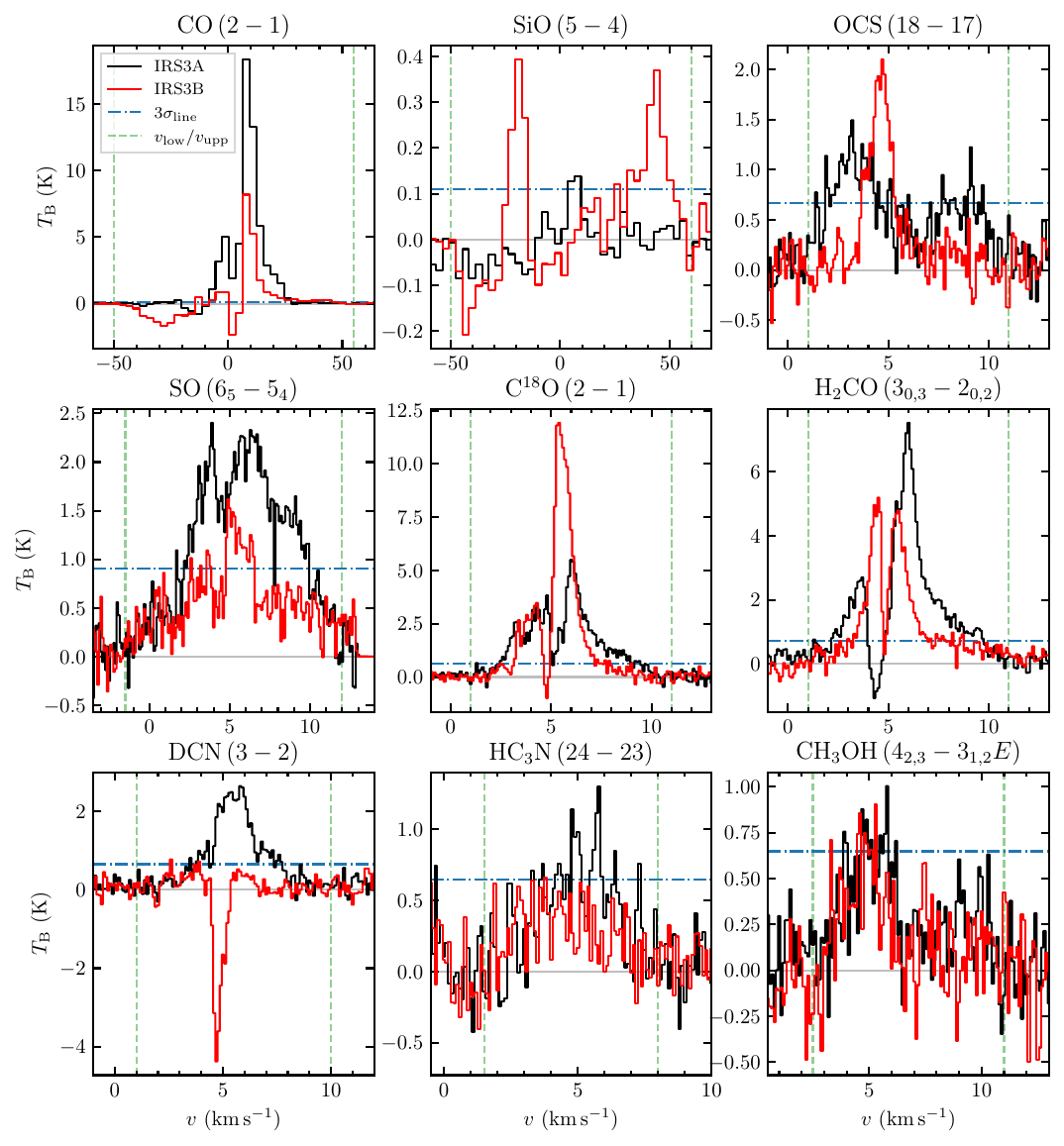}
\caption{Spectra toward the continuum peak positions of IRS3A (black) and IRS3B (red). The blue dash-dotted line marks the $3\sigma_\mathrm{line}$ level ($\sigma_\mathrm{line}$ is listed in Table \ref{tab:obs} for all transitions). The green vertical dashed lines mark the velocity integration range, $\varv_\mathrm{low}$ and $\varv_\mathrm{upp}$ (Table \ref{tab:obs}), used to compute the moment 0 maps (Fig. \ref{fig:moment0}).}
\label{fig:avg_spectra}
\end{figure*}

	Figure \ref{fig:avg_spectra} shows spectra extracted from the IRS3A and IRS3B continuum peak positions of all transitions (Table \ref{tab:obs}). The spatial morphology of all transitions within the FoV is presented in Fig. \ref{fig:moment0}.

\section{Outflow morphology of L1448N and L1448-mm}

\begin{figure}[!htb]
\centering
\includegraphics[]{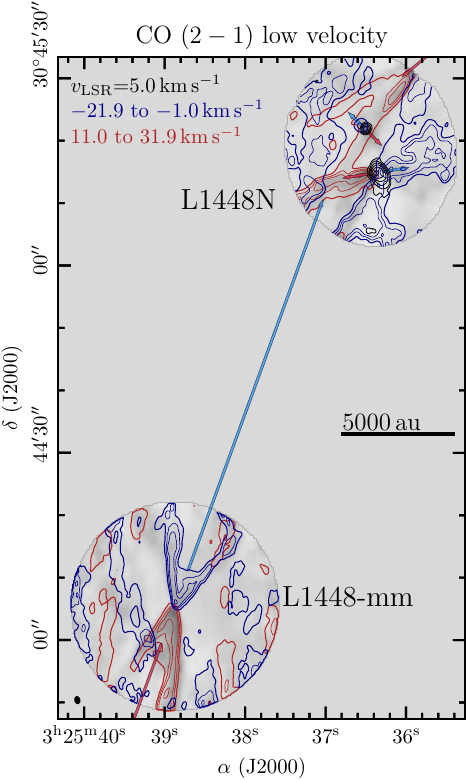}
\caption{The same as the left panel in Fig. \ref{fig:outflow}, but including the PRODIGE observations of the nearby L1448-mm protostar.}
\label{fig:outflow_largescale}
\end{figure}

	The two NOEMA pointings toward L1448N reveal extended CO and SiO emission that are not only caused by the IRS3A and IRS3B outflows, but there are also contributions by the nearby L1448-mm and IRS3C protostellar outflows. Figure \ref{fig:outflow_largescale} shows the low-velocity outflow components of CO ($2-1$) in red and blue contours of the L1448N mosaic (Fig. \ref{fig:outflow}) in comparison to the PRODIGE observations toward L1448-mm. A detailed analysis of the L1448-mm data will be presented in future work.

\section{Kinematic analysis of DCN, HC$_{3}$N, and CH$_{3}$OH}
\begin{figure}[!htb]
\centering
\includegraphics[]{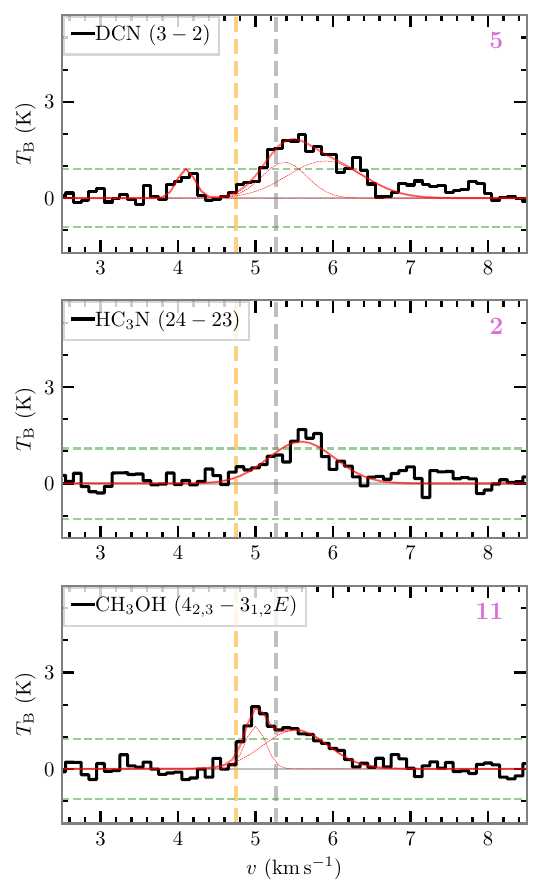}
\caption{Examples of observed spectra and Gaussian decomposition. The observed spectra are shown in black (top: DCN, middle: HC$_{3}$N, and bottom: CH$_{3}$OH) and are extracted from bridge positions (Fig. \ref{fig:bridgeoverview}) marked in the upper right. The total fit is indicated by the solid red line, while individual Gaussian velocity components are shown by the red dashed lines. The source velocity of IRS3A and IRS3B are indicated by the dashed grey and orange line, respectively. The line noise ($\pm$5$\sigma_\mathrm{line}$, Table \ref{tab:obs}) are indicated by horizontal dashed green lines.}
\label{fig:examplefit}
\end{figure}

	In Sect. \ref{sec:Gaussianfit}, we fit the observed spectra for DCN, HC$_{3}$N, and CH$_{3}$OH with up to three Gaussian velocity components. Example spectra at several positions (positions are marked in pink in Fig. \ref{fig:bridgeoverview}) along the gas bridge and corresponding multi-component Gaussian fits are presented in Fig. \ref{fig:examplefit}. 

\begin{figure*}[!htb]
\centering
\includegraphics[]{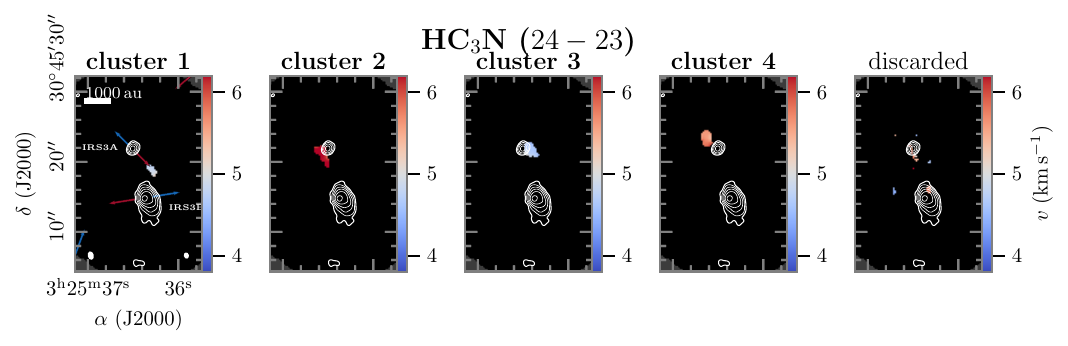}\\
\includegraphics[]{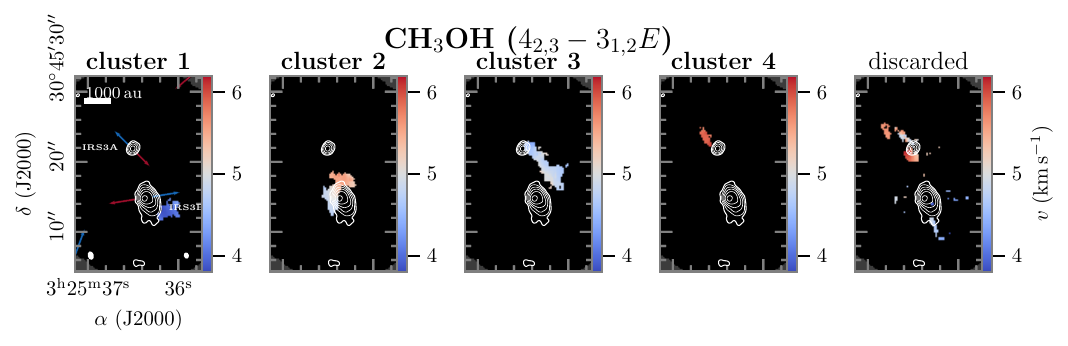}
\caption{The same as Fig. \ref{fig:clusters}, but for HC$_{3}$N (top panel) and CH$_{3}$OH (bottom panel).}
\label{fig:clusters_app}
\end{figure*}

	For DCN ($3-2$), the velocity component at $\approx$5.4\,km\,s$^{-1}$ is associated with cluster 2, while the component at $\approx$5.9\,km\,s$^{-1}$ is connected to the red part of the gas bridge (cluster 4, Fig. \ref{fig:clusters}). The weak emission peak at a velocity of 4.1\,km\,s$^{-1}$ is discarded in the clustering process with \texttt{DBSCAN}. For HC$_{3}$N ($24-23$) the component at $\approx$5.6\,km\,s$^{-1}$ is connected to the redshifted part of the gas bridge (cluster 4, Fig. \ref{fig:clusters_app}). For CH$_{3}$OH, the velocity at 5.5\,km\,s$^{-1}$ is associated with the envelope of IRS3B (cluster 2), while the velocity at narrower velocity component at 5.0\,km\,s$^{-1}$ is connected to the blueshifted side of the gas bridge (cluster 3, Fig. \ref{fig:clusters_app}).

\begin{table*}[!htb]
\caption{Kinematic properties of DCN ($3-2$), HC$_{3}$N ($24-23$), and CH$_{3}$OH ($4_{2,3}-3_{1,2}E$).}
\label{tab:dbscan}
\centering
\renewcommand{\arraystretch}{1.1}
\begin{tabular}{lr|rrr|rrr|r}
\hline\hline
 & Velocity & \multicolumn{3}{c|}{\texttt{DBSCAN input}} & \multicolumn{3}{c|}{\texttt{DBSCAN results}} & Global \\
 & components & data points & & & & clustered & discarded & velocity gradient \\
Transition & $n_{\mathrm{Gauss,max}}$ & (x,y,v) & \texttt{eps} & \texttt{minsamples} & cluster & \multicolumn{2}{c|}{data points} & (km\,s$^{-1}$\,pc$^{-1}$)\\
 \hline
DCN & 3 & 2856 & 0.30 & 95 & $n_\mathrm{cluster,tot} = 4$ & 1999 & 857 & \\
red bridge &&&&&cluster 4&647&& 52\\
blue bridge &&&&&cluster 3&542&& 77\\ \hline
HC$_{3}$N & 1 & 545 & 0.208 & 20 & $n_\mathrm{cluster,tot} = 4$ & 466 & 79 & \\
red bridge&&&&&cluster 4&126&& 116\\
blue bridge&&&&&cluster 1&61&& 131\\ \hline
CH$_{3}$OH & 2 & 1962 & 0.280 & 75 & $n_\mathrm{cluster,tot} = 4$ & 1500 & 462 & \\
red bridge&&&&&cluster 4&106&& 52\\
blue bridge&&&&&cluster 3&606&& 69\\
\hline
\end{tabular}
\tablefoot{$n_{\mathrm{Gauss,max}}$ is the maximum number of velocity components per spectrum in the Gaussian decomposition (Sect. \ref{sec:Gaussianfit}). The clustering method using \texttt{DBSCAN} is described in Sect. \ref{sec:clustering} and the global velocity gradient is the average gradient among the velocity gradient maps presented in Figs. \ref{fig:velocity_gradient}, \ref{fig:velocity_gradient_app1}, and \ref{fig:velocity_gradient_app2}.}
\end{table*}

\begin{figure*}[!htb]
\centering
\includegraphics[]{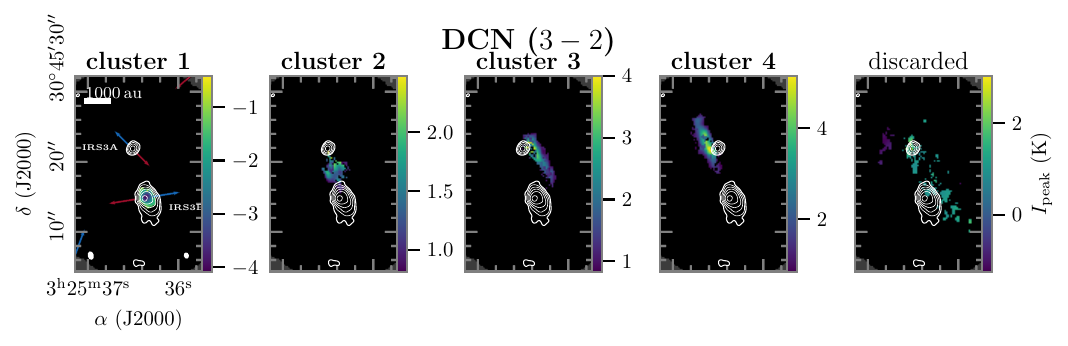}\\
\includegraphics[]{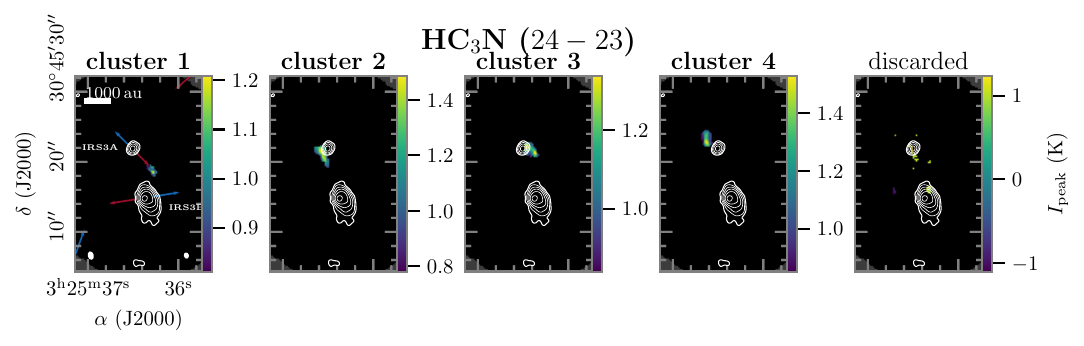}\\
\includegraphics[]{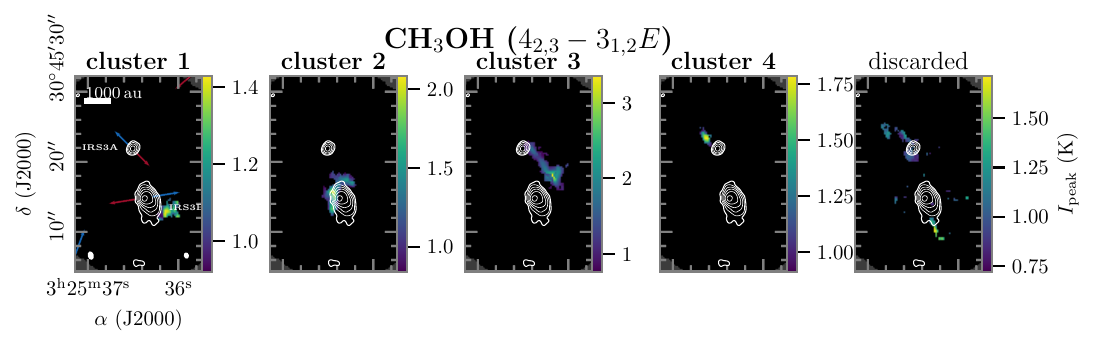}
\caption{The same as Fig. \ref{fig:clusters}, showing the corresponding peak intensity $I_\mathrm{peak}$ of the velocity clusters for DCN (top panel), HC$_{3}$N (middle panel) and CH$_{3}$OH (bottom panel).}
\label{fig:clusters_app_ampli}
\end{figure*}

\begin{figure*}[!htb]
\centering
\includegraphics[]{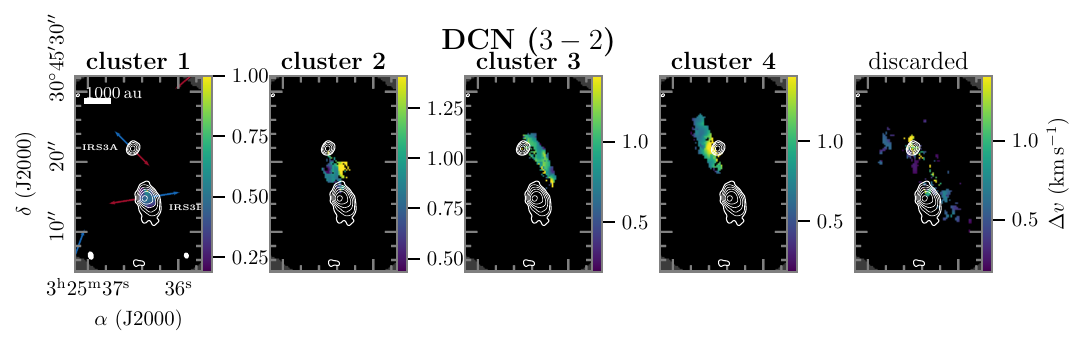}\\
\includegraphics[]{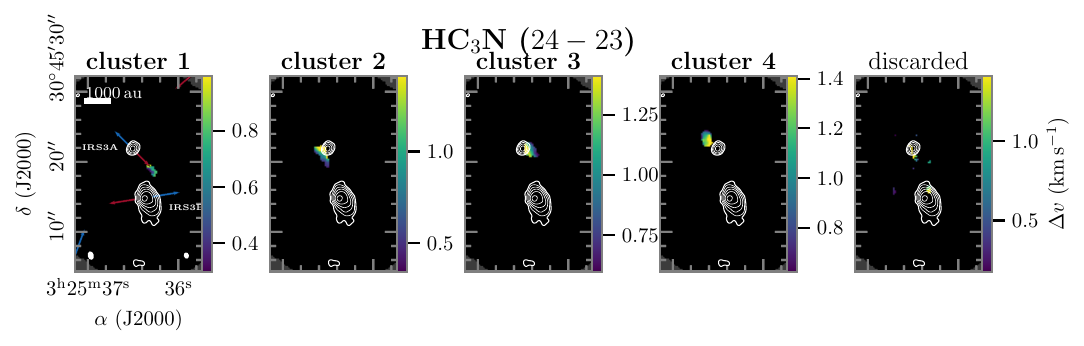}\\
\includegraphics[]{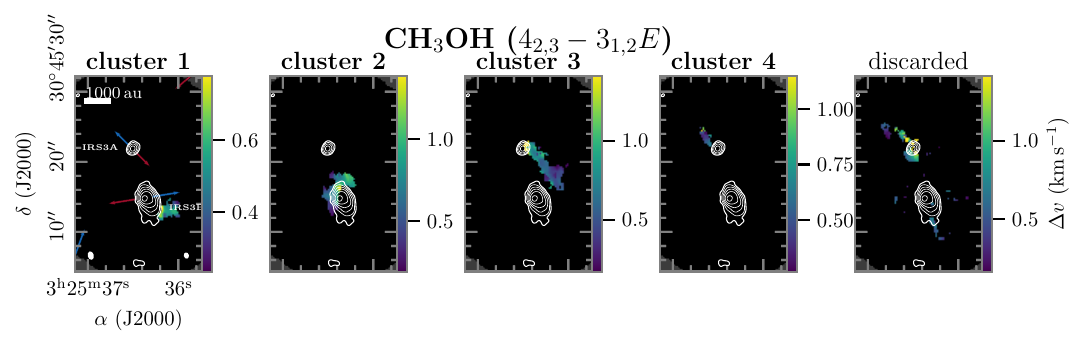}
\caption{The same as Fig. \ref{fig:clusters}, showing the FWHM line width $\Delta \varv$ of the velocity clusters for DCN (top panel), HC$_{3}$N (middle panel) and CH$_{3}$OH (bottom panel).}
\label{fig:clusters_app_width}
\end{figure*}

	The velocity components derived from the multi-component Gaussian modeling are clustered into their underlying physical origin using \texttt{DBSCAN} (Sect. \ref{sec:clustering}). While the velocity clusters of DCN are shown in Fig. \ref{fig:clusters}, the results for the HC$_{3}$N and CH$_{3}$OH are presented in Fig. \ref{fig:clusters_app}. The corresponding amplitude and line width maps are shown in Figs. \ref{fig:clusters_app_ampli} and \ref{fig:clusters_app_width}, respectively. 
	
\begin{figure*}[!htb]
\centering
\includegraphics[]{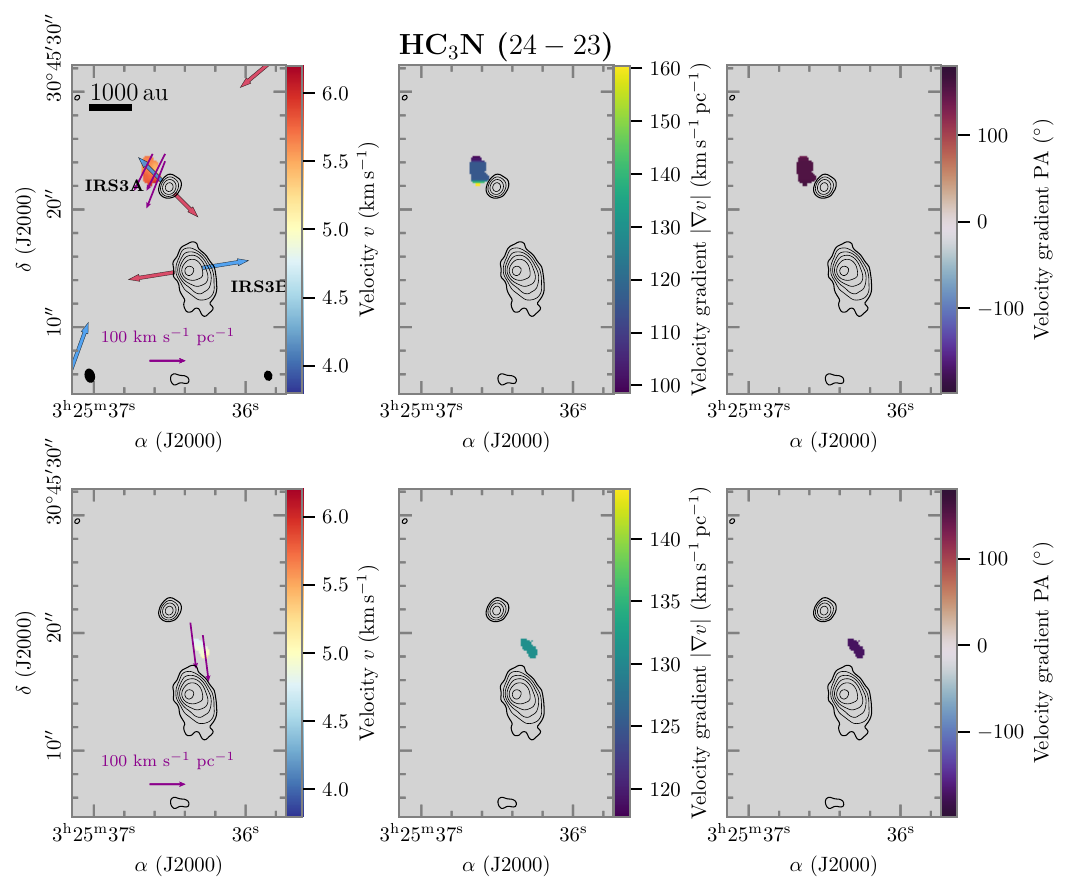}\\
\caption{The same as Fig. \ref{fig:velocity_gradient}, but for HC$_{3}$N ($24-23$).}
\label{fig:velocity_gradient_app1}
\end{figure*}

\begin{figure*}[!htb]
\centering
\includegraphics[]{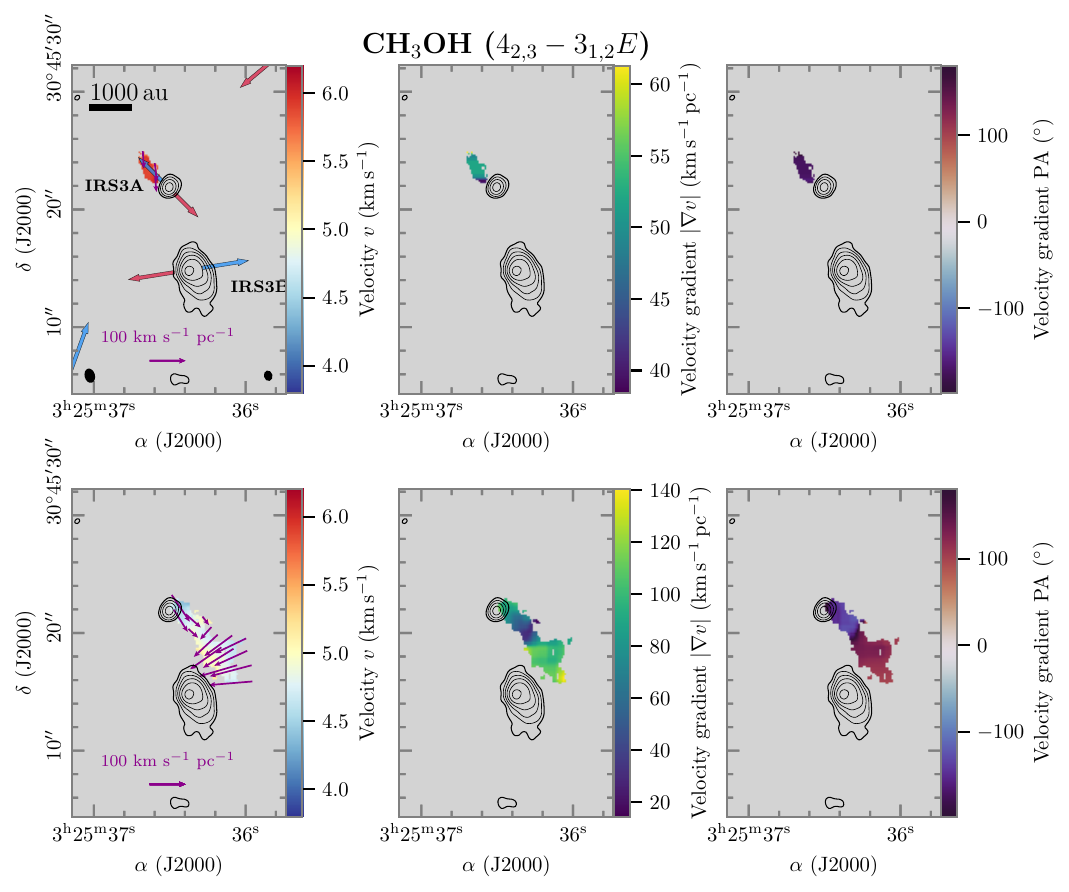}
\caption{The same as Fig. \ref{fig:velocity_gradient}, but for CH$_{3}$OH ($4_{2,3}-3_{1,2}E$).}
\label{fig:velocity_gradient_app2}
\end{figure*}

	The velocity gradients of the gas bridge are discussed in Sect. \ref{sec:velocitygradient}. The velocity gradient map of DCN is shown in Fig. \ref{fig:velocity_gradient}, while the results for HC$_{3}$N and CH$_{3}$OH are shown in Figs. \ref{fig:velocity_gradient_app1} and \ref{fig:velocity_gradient_app2}, respectively. The input and results from the clustering with \texttt{DBSCAN} and the global velocity gradients of each part of the gas bridge are summarized in Table \ref{tab:dbscan}.

\end{appendix}
\end{document}